\documentclass[aps,prb,showpacs,raggedbottom,nobalancelastpage,amssymb]{revtex4}
\usepackage[colorlinks=true,citecolor=blue]{hyperref}
\usepackage{amsmath}
\usepackage{amsfonts}
\usepackage{graphicx}
\usepackage{appendix}
\usepackage{dsfont}
\usepackage{amssymb}
\usepackage{bm}
\usepackage{color}
\usepackage{ulem}
\usepackage{soul}
\usepackage{natbib}
\usepackage{nccmath}
\usepackage{array}
\newcommand{\beq}{\begin{equation}}
\newcommand{\eneq}{\end{equation}}
\newcommand{\be}{\begin{equation}}
\newcommand{\ee}{\end{equation}}
\newcommand{\bea}{\begin{eqnarray}}
\newcommand{\eea}{\end{eqnarray}}

\usepackage{multirow}
\usepackage{wasysym}

\begin{document}
\title{ 
Current transport properties and phase diagram of a Kitaev chain with long-range pairing}
\author{Domenico Giuliano$^{(1,2)}$, Simone Paganelli$^{(3)}$, and Luca Lepori$^{(3,4)}$}
\affiliation{
$^{(1)}$ Dipartimento di Fisica, Universit\`a della Calabria Arcavacata di 
Rende I-87036, Cosenza, Italy \\
$^{(2)}$ I.N.F.N., Gruppo collegato di Cosenza, 
Arcavacata di Rende I-87036, Cosenza, Italy\\
$^{(3)}$ Dipartimento di Scienze Fisiche e Chimiche, Universit\`a dell' Aquila, via Vetoio, I-67010 Coppito-L'Aquila, Italy \\
$^{(4)}$ I.N.F.N., Laboratori Nazionali del Gran Sasso, Via G. Acitelli, 22, I-67100 Assergi (AQ), Italy }
\date{\today}

\begin{abstract} 

We describe a method to probe the quantum phase transition between   the short-range  topological phase and the long-range topological phase 
in the  superconducting Kitaev chain with long-range pairing, both exhibiting  subgap modes localized at the edges. 
The method relies on  the effects of the  finite mass of the subgap edge modes in the long-range regime
(which survives in the thermodynamic limit) on the single-particle scattering  coefficients 
through the chain connected to two normal leads. Specifically, we 
show that, when the leads are biased at a voltage $V$ with respect to the superconducting chain,     
the Fano factor is either zero (in the short-range correlated phase)  or $2e$ (in the long-range correlated phase).  As a result, we find that the 
Fano factor works as a directly measurable  quantity to probe the  quantum phase transition between the two phases. In addition, 
we note a remarkable  "critical fractionalization effect" in the Fano factor, which is exactly equal to $e$ along the quantum critical line. 
Finally, we note that a dual implementation of our proposed device makes it suitable as a generator of large-distance 
entangled two-particle states. 
 
\end{abstract}

\maketitle

\section{Introduction}
\label{intro}

The Kitaev chain provides a prototypical example of a one-dimensional (1D)  superconductive model with nontrivial topology \cite{kitaev}. 
The related phase diagram consists of two gapped phases, separated by a quantum critical point (QCP), at which the 
mass gap closes and the system undergoes a phase transition between a   topologically trivial and a topologically non trivial 
phase (TP). The hallmark of the emergence of the latter phase, which also constitutes its main point of interest, is the 
appearance of two unpaired real fermionic Majorana modes  $\gamma_L , \gamma_R$  (such that  
$\gamma_{L/R}^{\dagger} =  \gamma_{L/R}^{*} = \gamma_{L/R}  $), whose wave functions are localized at 
the endpoints of the chain \cite{stern_1,stern_2}. This phenomenon  is strictly related to the spontaneous breaking of the  
 ${\bf Z}_2$ fermion parity symmetry, due to the possibility of constructing a 
zero-energy Dirac mode $d = \frac{1}{2} [ \gamma_L + i \gamma_R ]$ which, on acting over a generic energy eigenstate 
$ | E \rangle$, changes its fermion parity, without changing its energy  \cite{sodsem}. 

Besides being interesting 
per se, the Kitaev Hamiltonian  also maps onto the 1D Ising model in a transverse magnetic field (TIM), by means of 
the standard Jordan-Wigner transformation (see for instance Ref.[\onlinecite{mussardo}]). Therefore, it also provides a 
way to exactly solve  the 1D TIM  by  diagonalizing a quadratic fermion Hamiltonian. Along the Jordan-Wigner transformation, the fermion parity 
${\bf Z}_2$ symmetry of the Kitaev Hamiltonian is traded for  the spin-parity ${\bf Z}_2$ symmetry in the TIM.  In particular, the topological phase in 
the former model corresponds to the ferromagnetic phase in the latter \cite{kitaev_2}. 

Due to the remarkable emergence of  a topological phase  and to the relevance of Majorana  modes as 
candidates for working as fault-tolerant quantum bits \cite{qubit}, the Kitaev chain has been largely studied in 
the last years. For instance, the effects on the Majorana 
modes of an additional electronic interaction along the chain  have been considered in Ref.[\onlinecite{oreg_alicea}], 
while the stability of a Majorana mode at the boundary 
of a Kitaev chain side coupled to an interacting normal wire has been discussed in Refs.[\onlinecite{fidkowski,giu_af_topo}]. 
Devices in which two Kitaev chains are connected to each other via a normal central region in an NSN-Josephson junction 
arrangement have been discussed as well in Ref.[\onlinecite{giu_af_josephson}], where  a particular focus has been put on the effects of 
the Majorana modes on the Josephson current flowing across the 
whole NSN junction when a fixed phase difference between the two superconductors is applied. Junctions of Kitaev chains 
have also been studied as a natural arena to realize and manipulate Majorana modes in a controlled way \cite{km_2}, as well 
as an equivalent model (via the Jordan-Wigner transformation) of junctions of quantum spin chains \cite{tsve_1,tsve_2,giu_x1,giu_x2}, or of suitably designed 
Josephson junction networks \cite{tsve_3,giu_x3}. On the experimental side, the Kitaev chain has been argued to provide an 
effective description of a superconducting proximity-induced 1D quantum wire with strong spin-orbit coupling and Zeeman effect
\cite{km_1,km_2}, which has accordingly been proposed as a feasible arena to experimentally look for emerging Majorana modes.  Indeed, 
following this scheme,  the Kitaev chain  has been experimentally realized in suitably  designed devices \cite{mourik2012,yazdani}. 

The above mentioned remarkable features of the Kitaev Hamiltonian have recently 
triggered  considerable interest in  generalizations of it, with possible additional novel 
phases. In this direction, a particularly interesting example is provided by 
the Kitaev chain with long-range pairing (LRK) \cite{lepori_1,lepori_3}, with generalizations to long-range 
hopping \cite{lepori_2,viy} and to Ising chains with long-range magnetic exchange strength  \cite{hauke,lepori_2,bradipo}, 
 as well as to higher-dimensional Hamiltonians \cite{viyuela2017,lgp} (it is also worth mentioning the possibility of 
 driving topological phase transitions by means of non-Abelian gauge potentials in optical  lattices \cite{lepori_x}). 
 The LRK is defined as a generalization 
of the Kitaev chain, with the pairing between particles at sites $i$ and $j$ decaying by a power-law 
function $\sim | i - j |^{-\alpha}$, $\alpha \geq 0$.  The standard Kitaev Hamiltonian, with 
pairing  involving only nearest-neighboring sites, is recovered in the limit $\alpha \to \infty$. 
On the experimental side, 
recent proposals have been put forward  to realize the LRK  and its generalizations by a Floquet engineering via an 
applied external AC field \cite{benito,zhao,nevado},  using neutral atoms loaded onto an optical lattice coupled to photonic modes 
\cite{rich,zol,sch_2,gop,sga,lukin2017,bettles2017}, or using Shiba bound states induced  in a chain of magnetic impurities  on the top  
of an s-wave superconductor  \cite{shiba_0,shiba_1}.

\begin{figure}
\centering
\includegraphics*[width=.4\linewidth]{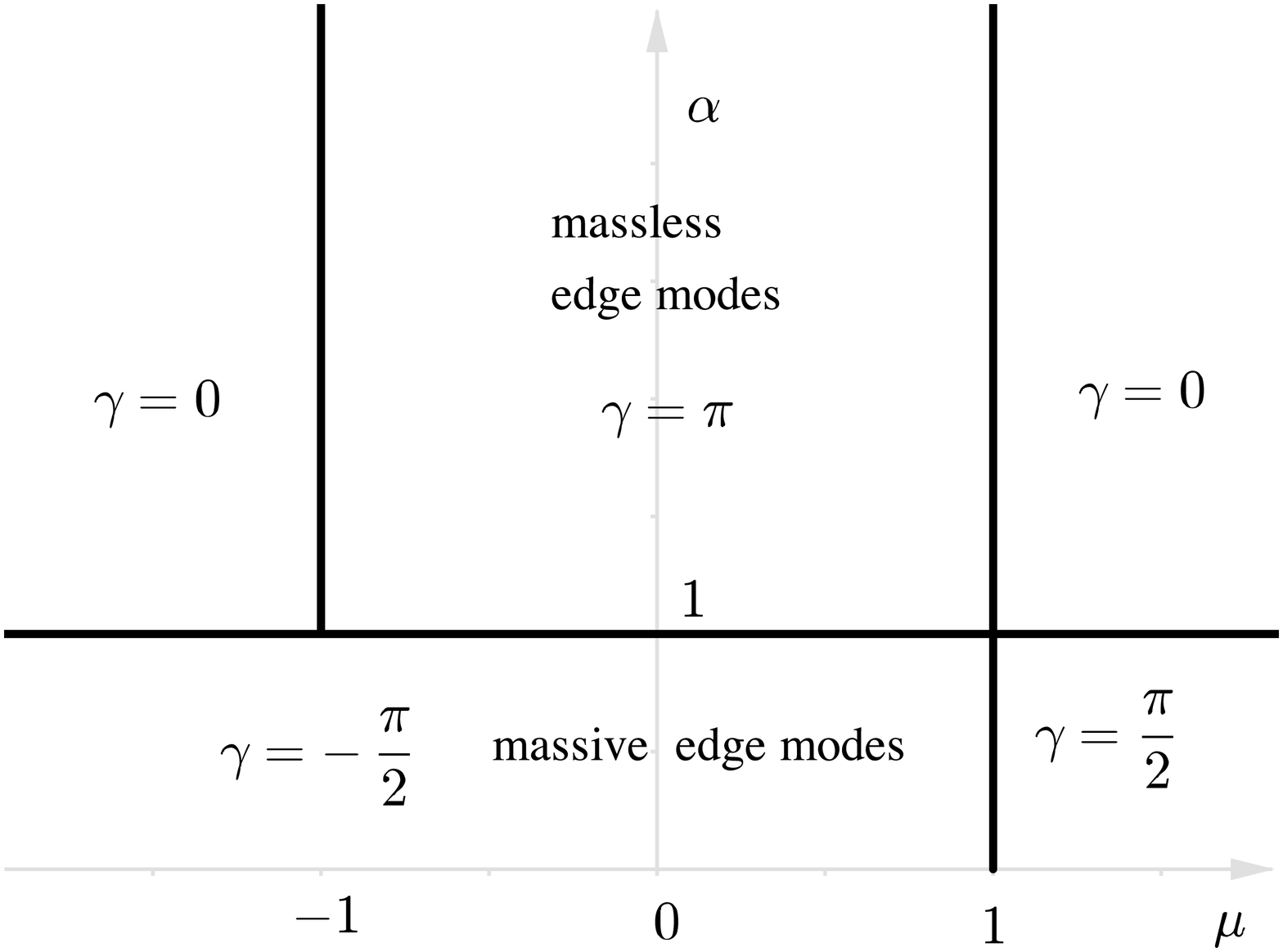}
\caption{Phase diagram of the LRK chain in the $\alpha$-$\mu$ plane: The various phases are characterized 
by different values of the Berry phase $\gamma$ (see main text for details)}
\label{phasediagram}
\end{figure}

As the LRK, either with short-range or with long-range pairing, is described by  a quadratic fermion Hamiltonian, 
it can be exactly solved within the standard approach to noninteracting fermion problems \cite{ripka}. This allowed for mapping out 
the whole phase diagram  in the  $\mu- \alpha$ plane   ($\mu$ being the chemical potential),   displayed in Fig. \ref{phasediagram}.
As a result,  two new  
phases have been found out \cite{lepori_1,lepori_2,viy,lepori_3,bradipo},   not continuously 
connected  to those characterizing the Kitaev model in the short range limit \cite{kitaev}. These phases have been also inferred,
not included in the  classification of topological insulators and superconductors valid in the short-range 
limit \cite{sr.1,sr.2,ludwig2009}. They are suggested and characterized by:

\begin{itemize}

 \item  The appearance of noninteger  winding numbers at $\alpha <1$ \cite{viy,bradipo}. 
 This property, as well as the ones described above, can be ascribed to the divergencies developing 
 in the quasiparticle spectrum for those values of $\alpha$ \cite{lepori_3,bradipo}. 
  Equivalently, a pertinently defined Berry phase $\gamma$ here takes the values $\gamma = \pm \, \pi /2$ (at $\mu \gtrless 1$),  
 different from the values $\gamma = 0, \pi$, which label
 the phases  of the Kitaev chain with short range  pairing.
 
 \item  The emergence of subgap modes localized at the edges of the open chain   in the phase at
 $\alpha <  1$ and $\mu <1$, as a remnant of the edge modes in the phase  at $\alpha > 1$ and $|\mu| <1$. 
 In the latter case, the structure of these subgap modes is qualitatively equivalent  to those in 
 the Kitaev model: Two real fermionic modes emerge, with wave functions 
 localized at the endpoints of the chain, which eventually evolve into the Majorana modes with vanishing mass in the infinite-chain limit. 
The emergence of the Majorana modes induces the spontaneous 
breaking of the fermion parity ${\bf Z}_2$ symmetry and it is the hallmark of the onset of a topological phase \cite{ludwig2009}
(this "short-range topological" phase will be denoted in the following as SRTP). \\
 On the contrary, for $\alpha < 1$, there are still
subgap modes with wave functions mostly localized around the endpoints of the chain, but with the corresponding wave function
overlap keeping finite, even in the infinite-chain limit. This corresponds to the onset,  in the same limit and at $\mu <1$, 
of a subgap mode with nonvanishing mass, determined by the hybridization of  
the two Majorana modes emerging  at $\alpha >1$ \cite{lepori_2,pachos2016}. Such phenomenon leads to a 
nondegenerate groundstate, thus restoring the ${\bf Z}_2$-symmetry. This restore
is by itself sufficient  to evidence the emergence  at $\mu <1$ and $\alpha < 1$ of
a phase not continuously connected to the topological phase of the standard (short-range) Kitaev model.

 \item  The violation of the area law for the Von Neumann entropy \cite{plenio_2}, also in the gapped regions,  
 as soon as $\alpha < 1$ \cite{lepori_1,lepori_3,ares}, and for every value of $\mu$.   On the contrary, the area 
 law is respected $\forall \, \alpha > 1$.  The mentioned violation has been shown \cite{ares} to be deeply related 
 to  the singularities in the quasiparticle spectrum,  originating the noninteger winding numbers.

 \end{itemize}

 The long-range correlated phase   at $\mu <1$ and $\alpha < 1$
is  also characterized by a suitably defined nontrivial LR topology
\cite{bradipo},  indeed reflecting in the massive subgap edge states.  For this reason, the same phase will be denoted as LRTP in the following. 
Finally, a  quantum phase transition at $\alpha = 1$, not first order (following the Ehrenfest scheme) \cite{lepori_1} and without any mass gap closure, 
can be inferred, 
also falling outside the standard schemes for the classification of 
quantum phases transitions, as presented in detail in, e.g., Ref.[\onlinecite{sachdev}].  
This result can be achieved mainly by other specific features, such as those listed above,  
and the divergence of the 
fidelity susceptibility \cite{zanardi2007} in $\alpha$ along the line $\alpha = 1$ \cite{lepori_5}.

\begin{figure}[ht]
\centering
\includegraphics*[width=.85\linewidth]{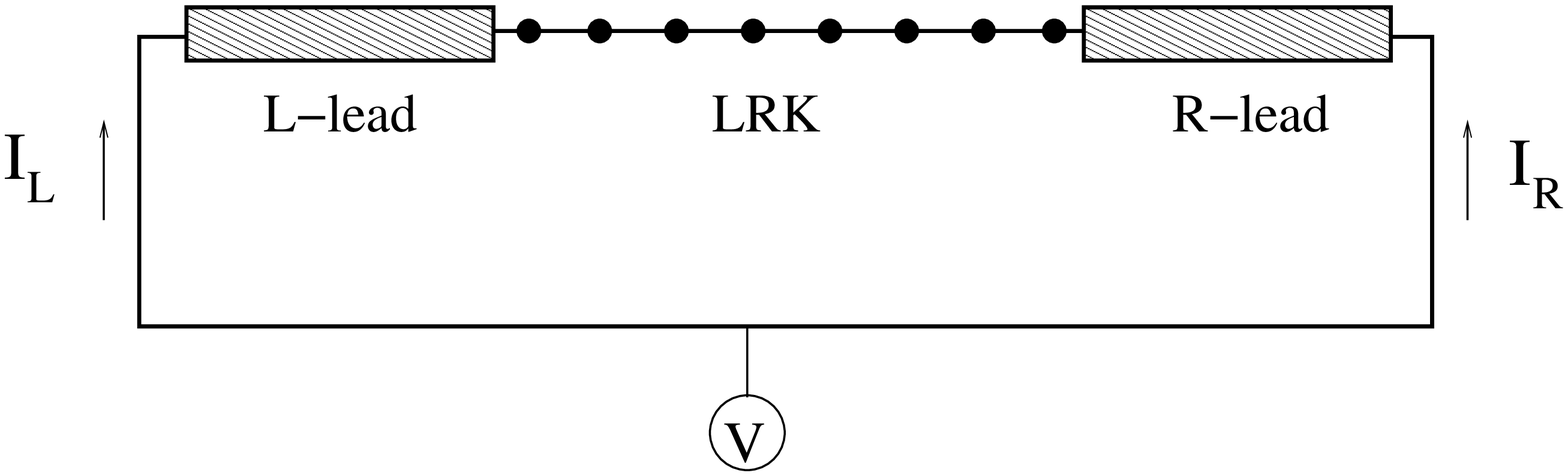}
\caption{Sketch of the NSN junction that we discuss in the paper. A LRK works as the central region of the junction; this region is 
connected to two normal leads, which can be biased at a finite voltage $V$ with respect to the LRK, thus 
making the currents $I_L$ and $I_R$, respectively, flow through the left-hand and  the right-hand SN interfaces. 
This is an adapted version of the device proposed in Ref.[\onlinecite{beenakker_1}].}
\label{device}
\end{figure}

In this paper we focus mainly on the  SRTP and LRTP,  characterized by the presence of subgap edge modes, and on the quantum
phase transition between the two of them. In particular, we discuss how the emergence of the massive subgap edge modes,   signaling  
 the onset of the LRTP at $\alpha <1$ and $\mu <1$, affects the single-particle scattering coefficients across the  LRK, when it is 
 connected to two normal leads, from which
particles and/or holes are injected  into the LRK and collected after scattering.  Specifically, we 
discuss an NSN device, in which the LRK is the central superconducting region, and the normal leads at its side can 
be biased to a finite voltage $V$ with respect to this region, so to make  an electric current flow across
the SN interfaces. In fact, our system, sketched in Fig. \ref{device}, can be regarded as an adapted version of 
the NSN junction studied in Ref.[\onlinecite{beenakker_1}] to discuss nonlocal Andreev reflection processes. To realize our  device, one needs   a solid-state
realization of the LRK Hamiltonian which can be recovered, for instance, as in Refs.[\onlinecite{shiba_0,shiba_1}], that 
is, by using helical Shiba states emerging at a chain of magnetic impurities deposited on top of an s-wave superconducting 
substrate. The level of control one may achieve in a system as such allows, in principle, to tunnel couple  the chain of magnetic 
impurities to normal contacts, which can be biased at a voltage $V$ and, at the same time, can be used to probe electric transport 
across the emerging LRK. Varying the superconducting coherence length of the host superconductor at a fixed length of the chain, one may also 
change the effective range of the induced pairing,   so to make  the system crossover from  an effectively short-range pairing regime 
to a long-range pairing one. Note that keeping  finite the length of the chain does not constitute an obstruction 
for the emergence of long-range physics, as described in detail in Ref.[\onlinecite{bradipo}]. 

 The key idea is that, when the leads are weakly coupled to 
the LRK in the SRTP, as $\ell$ grows, the low-energy (subgap) scattering processes across the NSN junction are expected to be fully determined by the 
two uncorrelated Majorana modes residing at the SN interfaces. As it happens with the Kitaev Hamiltonian,  this implies 
at the Fermi energy of the leads a strong suppression of all the scattering processes across each SN interface but the  (local) Andreev reflection (LAR),  
consisting in  the injection of a Cooper pair in the superconductive region,
via the absorption of a particle from the injecting lead and in the creation in the same lead of a counter propagating hole \cite{tinkham}, whose 
corresponding scattering coefficient flows to 1 \cite{giu_af_josephson,nava_1}. 
Instead, when the  system lies within the LRTP, the finite hybridization between the Majorana modes, yielding the massive 
subgap edge modes, is rather expected to lead to 
a full suppression of LAR, while keeping alive all the other scattering processes, including the remarkable nonlocal 
''crossed Andreev reflection'' (CAR) across the LRK, in which, differently from the LAR,  the hole  injected 
at one SN interface of the NSN system eventually emerges as a particle at the opposite interface \cite{beenakker_1}, 
with the corresponding scattering coefficients that keep finite at  the Fermi energy. 
 
Both LAR and CAR can make a finite current flowing across the SN interfaces when the leads are biased 
at a finite voltage $V$ with respect to the LRK. Nevertheless, as argued in Ref.[\onlinecite{beenakker_1}], a combined measurement 
of the current and of the zero-frequency current noise is able to discriminate whether it is the LAR or the CAR the process 
that is effective in supporting   the (low-$V$) current flow. 

Using this method, we look at the Fano factor, that is, at the 
ratio between the current noise and the current itself as $V \to 0$. We show that, whenever the current is supported by 
LAR (that is within the SRTP), the Fano factor flows to 0 as $V \to 0$, while, when the current is supported by 
CAR (that is within the LRTP), the Fano factor flows to $2 e$ in the same limit, keeping equal to $e$ exactly at
the phase transition line ($\alpha = 1$). Thus, on one hand we design a possible  experiment to 
discriminate between the two phases by means of a simple transport measurement. On the other hand, by 
considering a possible experiment  based on a process ''dual'' to CAR, in which one imagines to inject a Cooper pair from the superconductor into the leads 
as two outgoing particles,  we argue how the LRTP can be in principle used to create  pairs of distant (in real 
space),  highly-entangled particles.   To witness the reliability of the  combined measurements of 
current and noise to evidence the emergence of subgap modes in hybrid structures, it is worth stressing that 
it has been proposed to detect Majorana fermions at the edge of a vortex core in a chiral (two-dimensional) p-wave superconductor
\cite{bolech}, to probe Majorana modes at the interface between a superconductor and  (the surface of) a 
topological insulator \cite{law}, or in a Majorana fermion chain \cite{golub}, 
to measure Majorana fermions via transport through a quantum dot \cite{cao}.

In our case, while we acknowledge the difficulty of realizing our proposed NSN junction in a 
real solid-state device and of tuning $\alpha$ across the quantum critical line $\alpha = 1$, we 
believe that the experimental techniques mentioned above 
can make it possible to realize soon the junction in a controllable way. 

The paper is organized as follows:
\begin{itemize}
 \item In Sec. \ref{ki_lora} we introduce the model Hamiltonian for the NSN junction with the S-region realized by 
 the 1D LRK and review some basic features of the latter model. We then compute the single-particle/single-hole scattering amplitudes across the 
 LRK, as a function of the energy $E$ (measured with respect to the Fermi level of the leads)
 of the incoming particle/hole, paying particular attention to the $E \to 0$-limit.
 
\item In Sec. \ref{current_noise} we compute the current flowing through the leads when they are biased at a finite voltage $V$ with 
respect to the superconducting central region. We then compute the corresponding zero-frequency noise and the Fano factor within 
both the SRTP and the LRTP, highlighting the different behavior of the various physical quantities (currents and shot noise) in the two phases.

\item In Sec. \ref{crit_qua} we discuss the current, the  noise and  the Fano factor  across the quantum phase transition 
line at $\alpha = 1$.

\item In Sec, \ref{conclusions} we provide some concluding remarks and possible further developments of our work; then we
discuss  a possible solid-state implementation of the LRK. 

 \end{itemize}
Mathematical details concerning the calculations of the physically relevant quantities are provided in the appendices. In 
particular, in Appendix \ref{ssingle} we prove how our formalism (based on an extensive use of single-particle Green's functions to 
compute the various scattering amplitudes) is able to provide us back with the results of Ref.[\onlinecite{beenakker_1}] in the $\ell = 1$ limit.

\section{Model Hamiltonian and single-particle scattering processes}
\label{ki_lora}

In this section we introduce our main model Hamiltonian, discuss how we recover the scattering amplitudes within 
the imaginary time Green's function framework, discuss our results for the scattering coefficients, and compare them 
to the ones obtained within a simplified model adapted from Ref.[\onlinecite{beenakker_1}].

\subsection{The model Hamiltonian}
\label{ki_model}

 Throughout this paper, we consider the LRK as a reference model since. 
 As  studied and discussed in Refs.[\onlinecite{lepori_2,lepori_3}],  models characterized by a long-range hopping amplitude, as well, 
 do not give rise to qualitative modifications in the phase diagram. The corresponding LRK over an $\ell$-site lattice reads   \cite{lepori_1}  
\beq
H_{\rm LR} = - w \sum_{ j = 1}^{\ell - 1 } \{ d_j^\dagger d_{j+1} + d_{j+1}^\dagger d_j \} - 
\mu \sum_{ j = 1}^\ell d_j^\dagger d_j + \frac{\Delta}{2} \: \sum_{j = 1}^{\ell - 1 } \sum_{ r = 1}^{\ell - 1} \delta_r^{-\alpha} \{ d_j d_{j+r} + 
d_{j+r}^\dagger d_j^\dagger \}
\:\:\:\: , 
\label{klr.1}
\eneq
\noindent
with $\{ d_j , d_j^\dagger \}$ being lattice fermion operators and $\delta_r = | r |$ whenever $j+r \leq \ell$, otherwise $\delta_r = 0$.  
$H_{\rm LR}$ in Eq.(\ref{klr.1}) is a generalization of the (short-range)  Kitaev Hamiltonian 
\cite{kitaev}, to which it  reduces as $\alpha \to \infty$. In the following,  without any 
loss of generality, we conventionally choose the parameters of $H_{\rm LR}$ in analogy to what is done in 
\cite{lepori_1}: $\Delta = 2 w = 1$.

Incidentally, we note that, while for nearest-neighbor pairing, the Hamiltonian in Eq.(\ref{klr.1})
naturally emerges when considering the Jordan-Wigner representation of the one-dimensional Ising model in a transverse 
magnetic field, such a correspondence does not extend to the long-range Ising chain, due to the absence of cancellations between the Dirac 
strings in the Jordan-Wigner representation 
of the spin operators \cite{lepori_2};  indeed, this  makes it not  possible  
to solve the long-range Ising chain via the solution of the LRK.  

  In Fig. \ref{phasediagram}, we show the phase diagram of the Hamiltonian in Eq.(\ref{klr.1}) in the 
$\mu$ and $\alpha$ plane.  In the limit $\alpha \to \infty$, on varying $\mu$, two 
topological phase transitions are expected to take place at $\mu = \pm 1$, with the topologically nontrivial phase, where  
the Majorana modes  $\gamma_L$ and $\gamma_R$ appear  at the  endpoints of the open chain,  realized when $|\mu |< 1$. Within the same interval 
of values of $\mu$, but at $\alpha < 1$, the LRTP phase sets in, characterized by a finite hybridization energy $\epsilon_d$ 
between the two edge modes \cite{lepori_2,pachos2016}, which keeps finite in 
the thermodynamic limit and is typically accompanied by the onset  of a purely algebraic 
decay of the corresponding real-space wave functions \cite{lepori_1,lepori_2}. 

In order to discriminate between the two phases by means of an appropriate scattering experiment, we 
let particles and holes to be shot from the side reservoirs (leads) against the central region.
We model the normal leads by means of  noninteracting spinless fermion Hamiltonians 
${\cal H}_{\rm Lead} = \sum_{ X = L , R }  \: {\cal H}_X$, with 

\begin{eqnarray}
 {\cal H}_L &=& - J  \sum_{ j \leq -1 } \{ c_{ L , j}^\dagger c_{ L , j+1 } + c_{ L , j + 1}^\dagger c_{ L , j } \} - \mu' \sum_{ j \leq 0 }
 c_{ L , j}^\dagger c_{ L , j } \nonumber \\
  {\cal H}_R &=& - J \sum_{ j \geq \ell + 1 } \{ c_{ R , j}^\dagger c_{ R , j+1 } + c_{ R , j + 1}^\dagger c_{ R , j } \} - \mu' \sum_{ j \geq   \ell + 1} 
  c_{ R , j}^\dagger c_{ R , j }
  \:\:\:\: , 
  \label{eql.1}
\end{eqnarray}
\noindent
$c_{R / L , j }$ being single-fermion operators over each lead, $J$ and $\mu'$  the corresponding hopping amplitude and 
chemical potential, and, by convention, $j \leq 0$ for the left-hand lead and $j \geq \ell + 1$ for the right-hand lead. 
Finally, we  model the coupling between the central superconductive region and the leads by 
means of  the tunneling Hamiltonian 
\beq
{\cal H}_T = - t \{ c_{ L , 0}^\dagger d_1  + d_1^\dagger c_{ L , 0 } \} - 
t \{ c_{ R , \ell + 1}^\dagger d_\ell + d_\ell^\dagger c_{ R , \ell + 1 } \}
\:\:\:\: . 
\label{eql.7}
\eneq
\noindent
We remark that, using ${\cal H}_T$ as tunneling operator is equivalent to assuming purely local tunneling between 
S and the leads, despite intrinsically long-range nature of the 
correlations in S. This is justified   by making  
a weak coupling assumption between S and the leads, that is, $t / J \ll 1$.   Finally,  we note that, to simplify 
the derivation, we have chosen the lead Hamiltonian parameters in  Eq.(\ref{eql.1}), as well as 
the hopping amplitudes in Eq.(\ref{eql.7}),  $L-R$-symmetric, since  the results are qualitatively equivalent to 
what one gets  assuming different Hamiltonian parameters in ${\cal H}_R$ and ${\cal H}_L$ and/or different hopping amplitudes 
in ${\cal H}_T $. 

\subsection{The strategy to derive the scattering amplitudes}
\label{ki_strategy}

Scattering through the NSN junction is fully encoded the one-particle $S$ matrix  \cite{beenakker_1,nazarov}  $S(E)$, 
$E$ being the energy of the incoming particle from the leads at the beginning of the scattering process, measured with respect to the Fermi level of the leads. 
To set up the notation, in  the following we denote with $r_{X,X} ( E ) \: (\tilde{r}_{X,X} ( E ))$  the normal backscattering (NB) amplitude for a particle (hole) incoming from the
lead $X$, with  $a_{X,X} ( E ) \: (\tilde{a}_{X,X} ( E ))$ the LAR amplitude for a   particle (hole) incoming from the
lead $X$, with $t_{X,X'} ( E ) \: (\tilde{t}_{X,X'} ( E ))$ the normal transmission (NT) amplitude  for a particle (hole) incoming from 
the lead $X$ into the lead $X' \neq X$ and, finally, with $c_{X,X'} ( E ) \: (\tilde{c}_{X,X'} ( E ))$ the CAR amplitude  for a particle (hole) incoming from 
the lead $X$ into the lead $X'$. All the scattering amplitudes appear as entries of the $S$-matrix, which provides the outgoing state on pertinently acting onto 
the incoming state. In the low-energy limit in which one can assume that the particle- and hole-velocities are equal to each other and 
both equal to the Fermi velocity $v$, the $S$ matrix is given by 

\beq
S ( E ) = \left[ \begin{array}{cccc}
r_{L,L} ( E ) & \tilde{a}_{L,L} ( E ) & t_{L,R} ( E ) & \tilde{c}_{L,R} ( E ) \\
a_{L,L} ( E ) & \tilde{r}_{L,L} ( E ) & c_{L,R} ( E ) & \tilde{t}_{L,R} ( E ) \\
t_{R,L} ( E ) & \tilde{c}_{R,L} ( E ) & r_{R,R} ( E ) & \tilde{a}_{R,R} ( E ) \\
c_{R,L} ( E ) & \tilde{t}_{R,L} ( E ) & a_{R,R} ( E ) & \tilde{r}_{R,R} ( E ) 
                 \end{array} \right]
\:\:\:\: . 
\label{smatr.1}
\eneq
\noindent
Typically, for a quadratic Hamiltonian such as $H_{\rm LR}$, one may in principle construct $S(E)$ by 
solving the Bogoliubov-de Gennes (BDG) equations for the lattice energy eigenfunctions in the scattering basis, 
whose elements are labeled according to whether the incoming state corresponds to a particle or a hole, coming from 
the left-hand, or from the right-hand lead \cite{btk_1,btk_2}. While this procedure might in principle be 
applied to our system as well, by using the basis of scattering states that we review in 
Appendix \ref{scaboucon}, in fact, due to the long-range pairing term in $H_{\rm LR}$, the continuity conditions 
at the SN interfaces become quite hard to deal with.

For this reason, we resort to an alternative method \cite{bogoliubov}, based on 
the relation between the fully dressed Green's function of the NSN junction (which we exactly compute in 
Appendix \ref{construction}) and the $S$ matrix, which we review in detail in Appendix \ref{scattering_green}. 
Specifically, we first use the equations of motion, as implemented in Appendix \ref{scattering_green},
to prove that  $S (E )$ basically depends on the Green's function of the central region, ${\bf G}_{d ; ( j , j' ) } ( E )$, 
$j , j' \in \{ 1 , \ell \}$. Then, in order to optimize the numerical calculation,  we  resort to 
imaginary time formalism, eventually computing the Green's function over the imaginary axis in frequency space, ${\bf G}_{d ; ( j , j' ) } ( i \omega )$.
Finally, to recover the scattering amplitudes over the real axis, we analytically continue ${\bf G}_{d ; ( j , j' ) } ( i \omega )$ 
 for $\omega > 0$,  by means of the substitution $ i \omega \to E + i 0^+$. To 
perform the calculation, we numerically diagonalize $H_{\rm LR}$ at finite-$\ell$ for different values of $\mu$ and $\alpha$ and 
use the resulting eigenvalues and  eigenvectors to construct ${\bf G}_{d ; ( j , j' ) } ( i \omega )$. The relation between
$ S ( E ) $ and  ${\bf G}_{d ; ( j , j' ) } ( i \omega \to E + i 0^+)$ is readily derived by combining Eqs. (\ref{relfin}) of 
Appendix \ref{scattering_green} with Eqs. (\ref{sssc.2}, \ref{sssc.3}, \ref{ultimate.2}, \ref{ultimate.3}) of Appendix \ref{fully_dress}.

\subsection{Results}
\label{ki_results}

\begin{figure}[ht]
\centering
\includegraphics*[width=.85\linewidth]{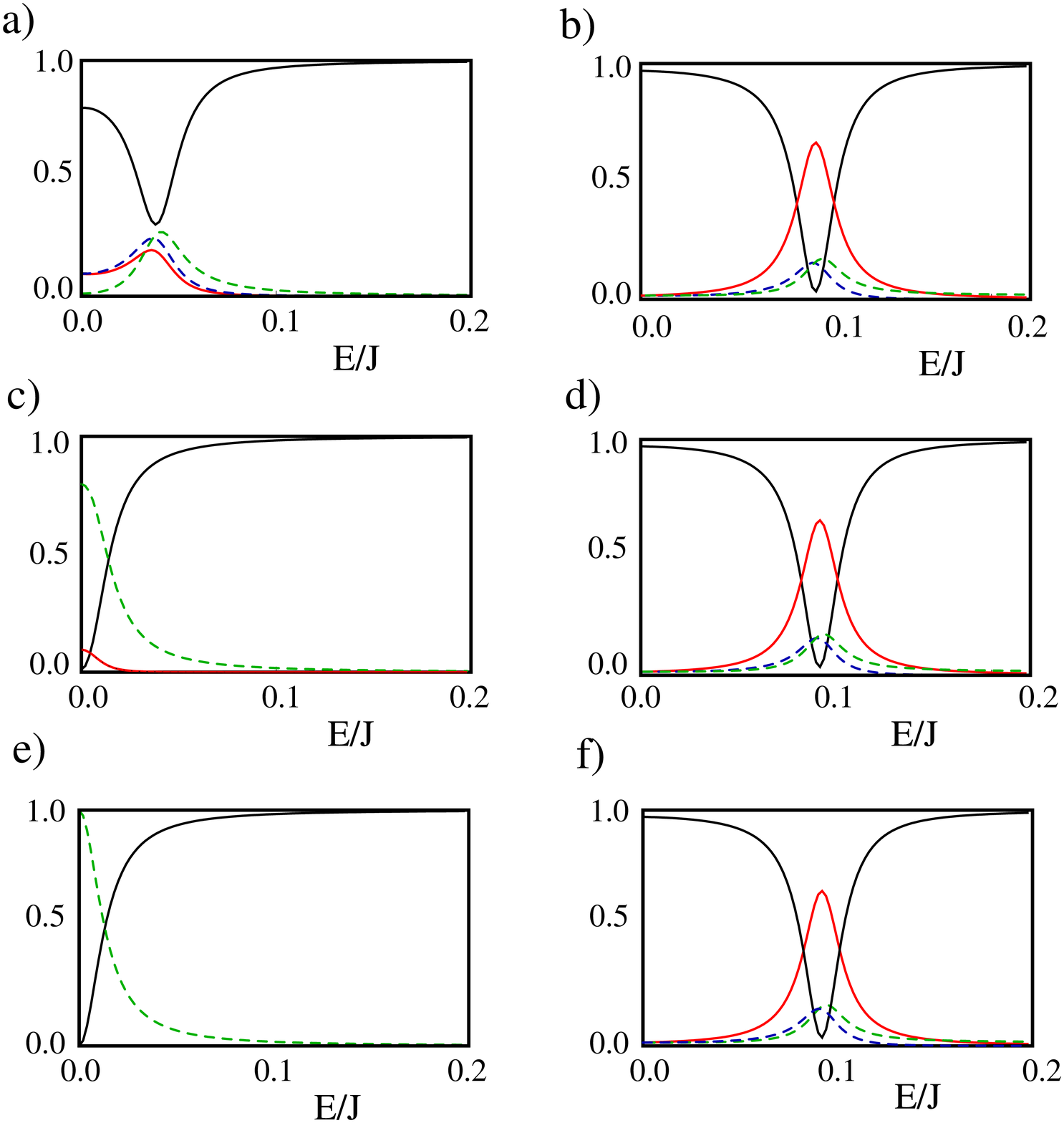}
\caption{ \\
 {\bf (a)}  Scattering coefficients $ R (E) $  (black solid line), $ A(E)$ (green dashed line), 
$  T (E) $ (red solid line) and $  C ( E ) $ (blue dashed line), calculated using $H_{\rm LR}$ in Eq. (\ref{klr.1}) as the Hamiltonian of
the central region, with $\Delta = 2, J= w = 1,\mu=0.25$ (SRTP), $\ell = 8$, and $\alpha = 10.5$. The 
Fermi momentum of the lead has been set at $k_F = \frac{\pi}{2}$ (half-filling), moreover the tunneling amplitudes between the leads and the 
central region have been symmetrically chosen to be $t /J = 0.2$. \\
{\bf (b)} Same as in panel {\bf a)}, but with $\alpha = 0.5$, that is deeply within the LRTP. \\ 
{\bf (c)} Same as in panel {\bf a)}, but with   $\ell = 16$ and $\alpha = 10.5$. \\ 
{\bf (d)} Same as in panel {\bf a)}, but with  $\ell = 16$ and $\alpha = 0.5$. \\  
{\bf (e)} Same as in panel {\bf a)}, but with $\ell = 26$ and $\alpha = 10.5$. \\  
{\bf (f)} Same as in panel {\bf a)}, but with  $\ell = 26$ and $\alpha = 0.5$.}
\label{scat_coef}
\end{figure} 
\begin{figure}[ht]
\centering
\includegraphics*[width=.85\linewidth]{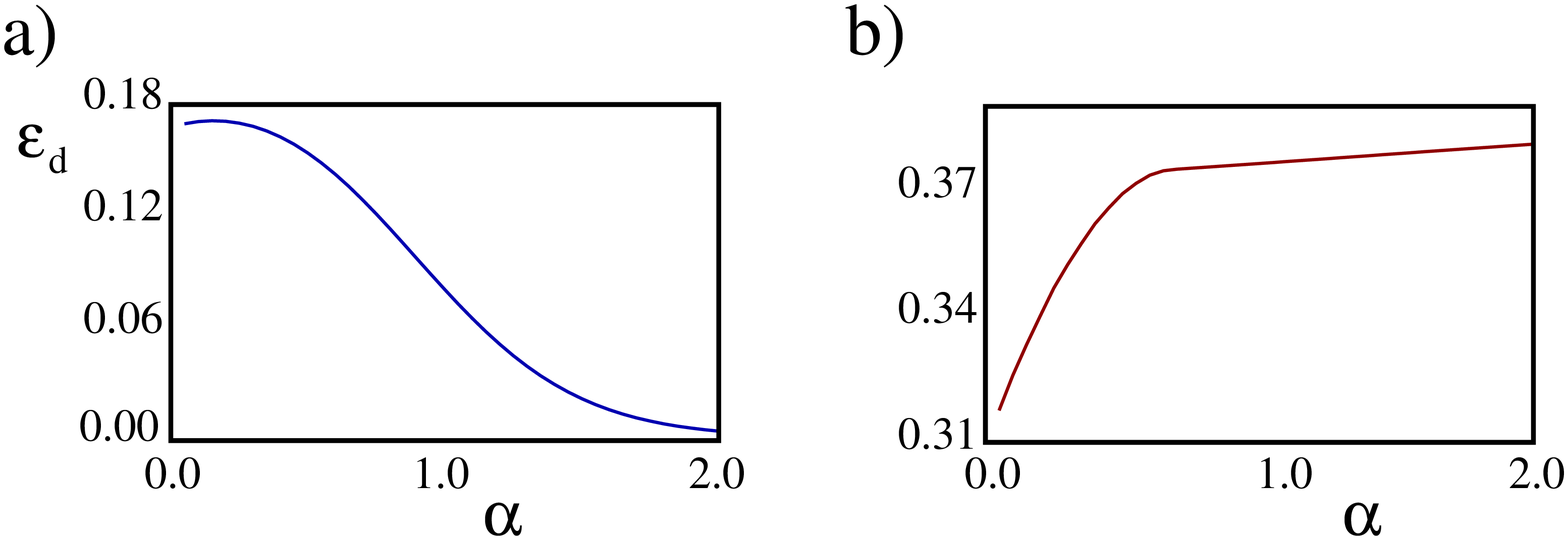} 
\caption{\\
{\bf (a)} Mass $\epsilon_d$ of the subgap mode as a function of $\alpha$ for the LRK described by the 
Hamiltonian in Eq.(\ref{klr.1}) with $\Delta = 2 , w = 1 , \mu=0.25$, and $\ell = 26$. \\ 
{\bf (b)}   Bulk mass gap for the same system, computed   assuming the same values of the  parameters as at point {\bf (a)}.}
\label{masses}
\end{figure} 

Because of the symmetries of our model Hamiltonian,  there are only four relevant scattering coefficients, respectively 
given by $ R ( E ) =  | r_{X,X} ( E ) |^2$ (NR),   $A ( E ) =  | a_{X,X} ( E ) |^2$ (LAR), $T ( E ) = | t_{X,X'} ( E )|^2$ (NT), 
and $C ( E ) = | c_{X,X'} ( E )|^2$ (CAR).

In  Fig.\ref{scat_coef} we draw 
the coefficients vs. $E$, for various values of $\ell$ and $\alpha$ (note that, as throughout all the paper, in drawing those
figures we assumed $w=J$ which, while simplifying the numerics, does not qualitatively affect the main results of 
the calculation). 
In all three cases ($0 \leq E \leq 0.2$ in units of $J$), $E$ 
lies within the gap, where there are no propagating modes.
It is therefore natural to attribute the features of the scattering coefficients to the emergence of the 
subgap modes and to their dynamics. To better spell out this, in Fig.\ref{masses} we display  the mass of the subgap edge modes and the bulk energy 
gap at $\ell = 26$ as a function of $\alpha$. 

 Within the SRTP  (Figs.\ref{scat_coef}{\bf (a)}, \ref{scat_coef}{\bf (c)}, \ref{scat_coef}{\bf (e)}), 
the overlap between the   subgap modes localized at the endpoints 
of the chain rapidly evolves to 0 as $\ell$ increases, starting from values comparable with the correlation length of the superconductor, 
thus turning them into the Majorana modes residing at the sides of the LRK in the SRTP \cite{lepori_1}.  
This implies a suppression of all the scattering processes at the interfaces with the leads but the LAR processes, which 
eventually take place with an amplitude of modulus 1 \cite{fidkowski,nava_1}. In fact, this is exactly what one infers by sequentially 
looking at Figs.\ref{scat_coef}{\bf (a)}, \ref{scat_coef}{\bf (c)}, and \ref{scat_coef}{\bf (e)} (drawn for different values of $\ell$): There the suppression 
of all the scattering processes except  $ A (E)$ is quite evident.  On the contrary,   when $\alpha = 0.5$, that is deeply within the LRTP,  the  overlap between the 
massive subgap modes keeps finite  even in the $\ell \to \infty$   limit \cite{lepori_1}. Accordingly,  
Figs.\ref{scat_coef}{\bf (b)},\ref{scat_coef}{\bf ( d)}, and \ref{scat_coef}{\bf ( f)} show a feature, corresponding to a 
sort of antiresonance in  $ R (E)$ and to a  resonance in all the other coefficients, at a
subgap energy scale which one naturally identifies with the   mass  $\epsilon_d$ of the subgap edge modes  \cite{lepori_1,lepori_2}.
This expectation is confirmed by the explicit calculation of $\epsilon_d$, yielding $\epsilon_d  \approx  0.087 $ if $\ell=6$ and
$\epsilon_d \approx 0.093$ if $\ell =16$ and $\ell =26$, corresponding with good accuracy to the location in $E$ of the (anti)resonances in 
Figs.\ref{scat_coef}{\bf (b)}, \ref{scat_coef}{\bf (d)}, and  \ref{scat_coef}{\bf (f)}.

In the two phases at $\alpha >1$ and $|\mu|>1$ and at $\alpha <1$ and $\mu>1$, where no subgap edge modes are present at all, no resonance is visible.
Moreover CAR is extremely suppressed at every $E$ in the range examined, as  expected, as well the NT and the LAR in the limit $E \to 0$, in which 
same limit one eventually obtains  $R(E) \to 1$.  As we discuss in the following, 
this behavior deeply affects the dependence on $E$ of the currents as a voltage $V$ is applied to the leads.

\subsection{Comparison with an effective single-site model}
\label{ki_effectivedge}
 
To double check our interpretation of Fig.\ref{scat_coef}  in relations to the subgap excitation content
within the SRTP and the LRTP, we  refer to a simplified single-site model for the central region of the NSN juction, which we
review in detail in Appendix \ref{ssingle} and that is   just an adapted version of the model   discussed in Ref.[\onlinecite{beenakker_1}], 
yet able to catch the relevant subgap physics of the LRK. 

More specifically, following  Ref.[\onlinecite{beenakker_1}],  we
consider the Hamiltonian for the central region in the form 
\beq
 {\cal H}_d = i \, \epsilon_d \, \gamma_L \gamma_R
\;\;\;\; ,
\label{es.1text}
\eneq
\noindent
with $\gamma_L$ and $\gamma_R$ being real Majorana modes, respectively coupled to the left-hand and to the 
right-hand lead, so that   the tunneling Hamiltonian ${\cal H}_T$ accordingly reduces to 
\beq
{\cal H}_T = - t [ c_{L , 0 }^\dagger  - c_{L , 0 } - i  ( c_{R , \ell + 1 }^\dagger - c_{R , \ell + 1 }  )  ] d 
- t \, d^\dagger [ c_{L ,  0 } - c_{L , 0 }^\dagger + i ( c_{ R , \ell + 1 } - c_{R , \ell + 1 }^\dagger  ) ] 
\:\:\:\: , 
\label{es.5}
\eneq
with $d = \frac{1}{2} [ \gamma_L + i \gamma_R ]$.

\begin{figure}[ht]
\centering
\includegraphics*[width=.85\linewidth]{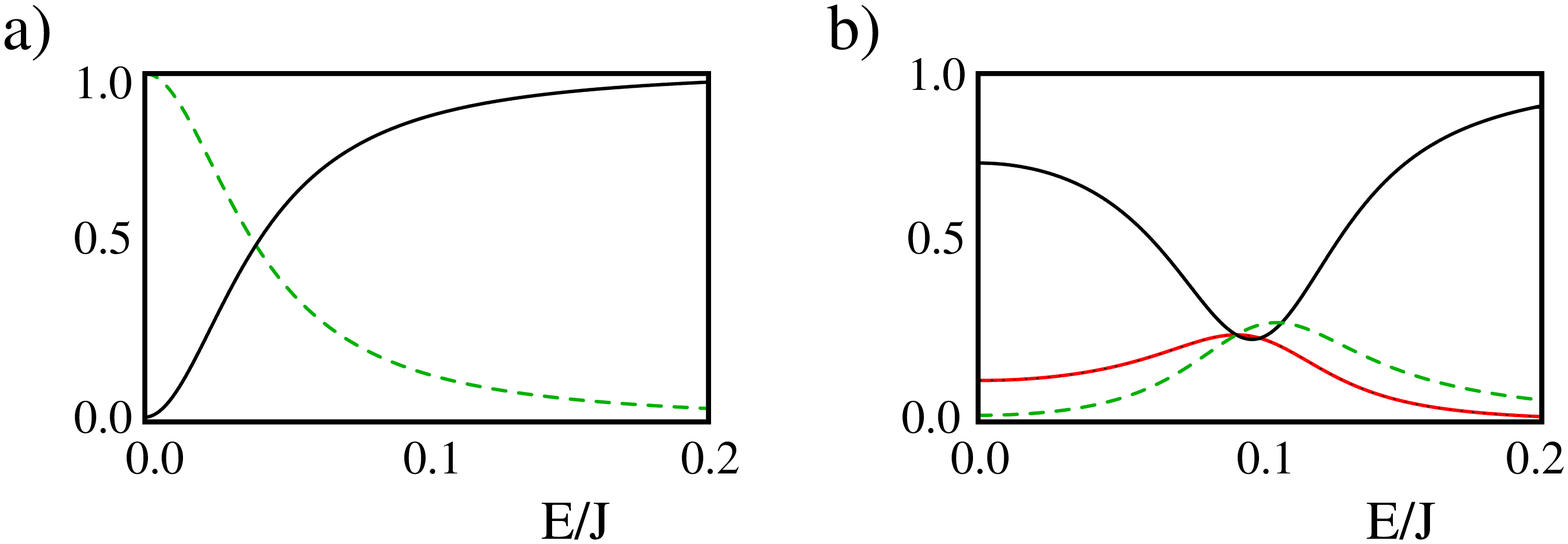}
\caption{\\
{\bf (a)}  Scattering coefficients $ R (E) $  (black solid line), $ A(E)$ (green dashed line), 
$  T (E) $ (red solid line), and $  C ( E ) $ (blue dashed line),    calculated using 
the effective model Hamiltonian  ${\cal H}_d + {\cal H}_T$, with $J=1, t/J=0.1, k_F = \frac{\pi}{2}$,   and $\epsilon_d=0$. 
Neither the red solid, nor the blue dashed line appear in the plot, due to 
the extremely low values taken by the corresponding coefficients. The Andreev reflection coefficient saturates to 1 as $E \to 0$, consistently 
with the fact that the Majorana modes are decoupled from each other as $\epsilon_d \to 0$, while the normal 
reflection coefficient goes to 0 in the same limit.\\
{\bf (b)} Same as in panel {\bf (a)}, but with $\epsilon_d = 0.1$, to mimic the case of long-range pairing   ($\epsilon_d = 0.1$ at $\alpha \approx 0.85$ if $\ell = 26$, see Fig.\ref{masses} ).
Note that the red solid and the blue dashed line collapsed onto each other \cite{beenakker_1}.}
\label{finite}
\end{figure}

In Fig.\ref{finite}, we 
plot the scattering coefficients versus the energy $E$ of the incoming particle/hole, computed using the Hamiltonian ${\cal H}_d + {\cal H}_T$. 
The various parameters in the same Hamiltonian are chosen equal to the the ones 
that we used to draw Fig.\ref{scat_coef}, included the parameter $\epsilon_d$, 
which we set at the appropriate value by using the numerical data displayed  in Fig.\ref{masses}{\bf a)}. 
One clearly sees that the plots in Fig.\ref{finite} behave qualitatively similarly to 
the corresponding ones derived  for the actual system inFig.\ref{scat_coef}. This result 
enforces our interpretation of the resonance-like features in the scattering coefficients reported in the latter figures 
as an evidence for the existence of a finite-mass subgap edge mode. 

To summarize the discussion of this section, we provided an evidence that one can infer the emergence of a subgap edge mode in a
length-$\ell$ Kitaev chain  with long-range pairing by just looking at the dependence on $E$  of the scattering amplitudes across the 
chain. At finite energy  (finite $\ell$ and/or $\alpha <1$), the subgap modes are evidenced as resonances/antiresonances in the scattering coefficients  
at energies below the bulk superconducting gap of the chain  and, as we are going to discuss in the following, they have important consequences 
for the transport properties of the whole NSN junctions.  Therefore, by looking at these properties, one expects to be able to directly probe 
the existence of a subgap edge mode in the system and whether it is locked at zero energy in the thermodynamic limit, as it happens in the 
SRTP, or at a finite  energy $\epsilon_d$, as it rather happens in the LRTP. 

\section{Current noise, Fano factor and scattering coefficients close to the Fermi surface} 
\label{current_noise}

The scattering coefficients we compute above determine also the electric transport 
properties across the system, when the normal leads are biased at a voltage $V$ with respect to the
central superconducting region.  As we show in this section,  the ratio between the total zero-frequency shot noise 
 in the current flowing through the system and the current itself (the Fano factor) allows for 
  discriminating between the SRTP and the LRTP.

To calculate the current, we follow a ''Landauer-Buttiker like'' approach \cite{nazarov}, 
by basically assuming that particles and holes are shot into the junction from thermal reservoirs at temperature 
$T$ (which we eventually send to 0,   so to recover the shot-noise regime, that is, $k_B T \ll eV$, with $k_B$ being
the Boltzmann constant). 

The main derivation of the formulas for the currents and for the current correlations is summarized in 
Appendix \ref{transport}, where we also show that the current flowing across the left- and right-hand lead, 
($I_L$ and $I_R$, respectively) is given by $I_L = - I_R = I$, with 
\beq
I = \frac{2 e}{2 \pi} \: \int_0^\infty \: d E \: \{ f (-  E - eV ) - f ( - E + e V ) \}
 \: \{  | a_{L,L} ( E )|^2  + | c_{L,R} ( E ) |^2 \}  
\:\:\:\:,
\label{transp.1}
\eneq
\noindent
$f ( E )$ being the Fermi distribution function at the temperature $T$ of the thermal reservoirs,   
while $a_{L,L} ( E )$ and $c_{L,R } ( E )$ are  the amplitudes for the (local and crossed) Andreev reflection, 
defined in Sec. \ref{ki_strategy}.  At zero temperature (and, more generally, in the shot-noise regime), Eq.(\ref{transp.1}) becomes 
\beq
I = \frac{2 e}{2 \pi} \: \int_0^{eV}  \: d E \: 
 \: \{ |a_{L,L} ( E ) |^2 + | c_{L,R} ( E ) |^2 \}  
\:\:\:\:.
\label{transp.2}
\eneq
\noindent
Equations \eqref{transp.1} and \eqref{transp.2} clearly show that a nonzero net current may flow toward the leads
even for $eV$ smaller than the superconducting gap of the central region, provided that  either the 
LAR amplitudes, or the CAR amplitudes (or both of them) keep 
different from zero close to the Fermi energy. Incidentally, this implies a strong suppression of the 
subgap electric transport within the non-topological phase at $\mu >1$ and $\alpha >1$, where, as described in Sec. \ref{ki_results}, all the scattering amplitudes but the 
ones corresponding to normal reflection processes go to zero when approaching the Fermi level,   $E \to 0$ . The same suppression occurs at $\alpha <1$
if $\mu >1$ and no massive subgap modes are present. 

As a special case of Eqs.(\ref{transp.1}) and (\ref{transp.2}), one may consider the limit of zero CAR  
amplitude, $c_{L , R } ( E \to 0 ) \to 0$. When accompanied by a suppression of the normal 
transmission amplitude as well [$t_{L , R} ( E \to 0 ) \to 0$], this limiting situation mimics what happens at 
a single NS interface, where the finite-temperature (zero-temperature)  current is given by 
Eq.(\ref{transp.1}) [Eq.(\ref{transp.2})], with $c_{L , R } (E) = 0$ \cite{btk_1,btk_2}. In the specific case of a 
LRK within the SRTP at $\alpha>1$, the emergence 
of the Majorana mode suppresses all the backscattering processes, except the LAR. 
This partial suppression makes the DC conductance $G$, associated to the current transport through the normal region,  
reach the maximum consistent with unitarity constraint  with an elementary carrier 
charge $e^* = 2 e $, that is, $G = \frac{2 e^2}{2 \pi}$ \cite{fidkowski,giu_af_topo}. 

In our specific case, as we see from the plots of the previous section within the SRTP and the LRTP, either  $ a_{L,L} ( E  )$ or 
$ c_{L,R} ( E ) $ (or both) keep nonvanishing as $E \to 0$, which makes it hard to disentangle, from a
measurement yielding a finite $I$ at low values of $e  V$, whether to attribute it to LAR or to CAR and, 
accordingly, whether to attribute a finite value of $I$ to the onset of the SRTP or of the LRTP.  Therefore, in order to define an 
 experimental mean to distinguish the two phases from each other via transport measurements, we 
follow the  approach of  Refs.[\onlinecite{bolech,law,beenakker_1}] and consider the zero-frequency shot noise  associated to $I$.  

In order to  evidence the $L-R$ symmetry in the formula for the shot noise, 
we use a symmetrized version of the operator associated to $I$,  given by $ J = \frac{1}{2} \{ J_j - J_{j'} \}$, with $j$ belonging to lead 
$L$ and $j'$ to lead $R$ and 
\beq
J_j = - i e J \{ c_j^\dagger c_{j+1} - c_{j+1}^\dagger c_j \} \, .
\label{cur.1_text}
\eneq
For $eV \gg k_B T$, we obtain 
\beq
I = \langle J ( t ) \rangle =  \frac{ e}{2 \pi} \: \int_0^{eV}  \: d E \: 
 \: \{ |a_{L,L} ( E ) |^2 + | a_{R,R} ( E ) |^2 +  |c_{L,R} ( E ) |^2 + | c_{R,L} ( E ) |^2\}
 \label{transp.3}  
\:\:\:\:,
\eneq
with $\langle \ldots \rangle$ denoting the thermal average with respect to the lead distribution functions. 
  On applying the definition of the shot noise to our specific system, we find that   
the zero-frequency shot noise  associated to $J$ at a voltage bias $V$, ${\cal P}( 0 , V )$   is given by   
\beq
{\cal P}( 0  , V )  =  \frac{1}{4} \{ {\cal P}_{L,L} ( 0 , V ) + {\cal P}_{R,R} ( 0, V  ) - { \cal P}_{L,R} ( 0 , V ) - {\cal P}_{R,L} ( 0, V  ) \} 
\;\;\;\; ,
\label{transp.4}
\eneq
\noindent
with ${\cal P}_{L,L} ( 0  )$, $ {\cal P}_{R,R} ( 0  )$ , ${ \cal P}_{L,R} ( 0  )$,  $ {\cal P}_{R,L} ( 0 ) $,
defined  in Eqs.(\ref{cur.17},\ref{cur.17bis},\ref{cur.19},\ref{cur.19bis}) of Appendix \ref{noise} as 
 combinations of the scattering amplitudes appearing in the $S(E)$ matrix in Eq.(\ref{smatr.1})   (for notational 
 simplicity, in the following we will not explicitly show the dependence of the shot noise on the voltage $V$).   

The key quantity we 
now look at is the Fano factor, defined as ${\cal P} ( 0  ) / I$. Within the SRTP, 
in the $e V\to 0$ limit, all the scattering coefficients but the LAR one are suppressed. This makes the SN junctions
behave as a perfect conductor, with conductance equal to  $G = \frac{2 e^2}{2 \pi}$ \cite{fidkowski,giu_af_topo} and
with a corresponding suppression of the zero-frequency shot-noise; therefore the Fano factor  is expected to flow to 0 as 
$e \,V \to 0$. On the contrary, within the LRTP, we infer from the plots in  
Figs.\ref{scat_coef}{\bf (b)},  \ref{scat_coef}{\bf ( d)}, \ref{scat_coef}{\bf (f)} that, though reduced, both the CAR and the 
LAR coefficients keep finite as $e V \to 0$, due to the emergence of the finite-energy massive subgap modes. In addition, 
also the NR and NT scattering coefficients keep finite as well when $e V \to 0$. While this is already enough to 
 expect a nonzero Fano factor as $e V \to 0$, we also note that, in the limit in which the leads are 
weakly coupled to the LRK (that is, at small values of $t / J$), the physical processes supporting the current 
transport across the NSN junction  become rare events. These processes are exactly the LAR and the CAR,
plus complementary processes obtained from the symmetries of the $S$ matrix. 
Since  the net charge of the elementary charge carrier is $e^* = 2 e$ (that is, the charge of 
a Cooper pair), we expect on one hand that the zero-frequency shot noise  becomes Poissonian, on the other hand that, 
as $e V \to 0$, the Fano factor converges to $e^*$, at least for large enough values of $\ell$. 

\begin{figure}[ht]
\centering
\includegraphics*[width=.8\linewidth]{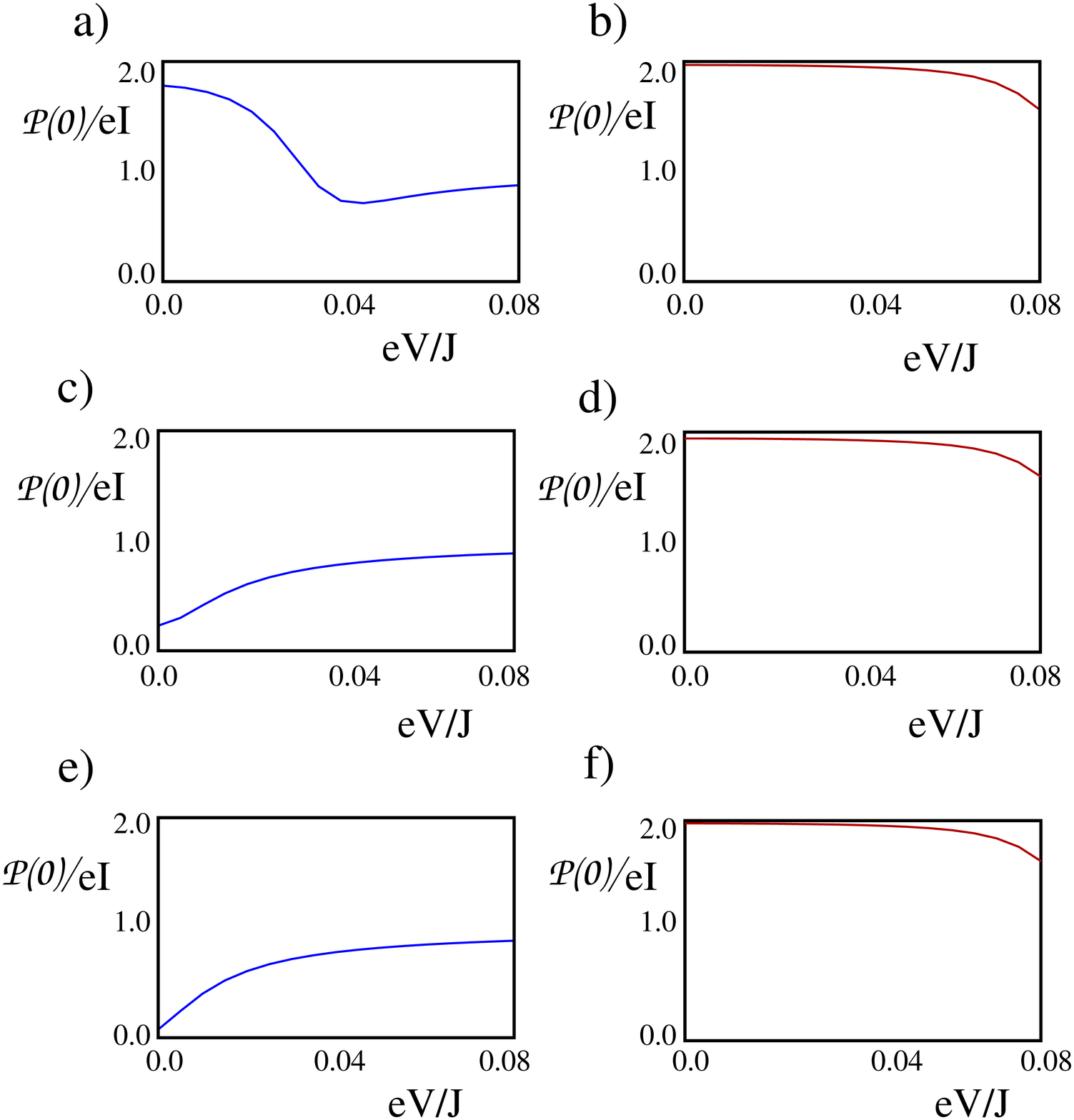}
\caption{{\bf (a)}  The Fano factor   ${\cal P} ( 0  ) / I e$ computed for the LRK
Hamiltonian with   $\Delta = 2 w = 2,\mu=0.25$, $J=w$, and $\ell = 8$ and $\alpha = 10.5$ (SRTP). The  
Fermi momentum of the leads has been set at $k_F = \frac{\pi}{2}$ (half filling),  while the tunneling amplitudes between the leads and the 
central region have been symmetrically chosen to be $t / J= 0.1$. \\
{\bf (b)} Same as in panel {\bf (a)}, but with $\alpha = 0.5$ (LRTP). \\
 {\bf (c)} Same as in panel {\bf (a)}, but with $\ell = 16$ and $\alpha = 10.5$. \\
{\bf (d)} Same as in panel {\bf (a)}, but with $\ell = 16$ and $\alpha = 0.5$. \\
 {\bf (e)}   Same as in panel {\bf (a)}, but with  $\ell = 26$  and $\alpha = 10.5$. \\
{\bf (f)} Same as in panel {\bf (a)}, but with  $\ell = 26$  and $\alpha = 0.5$.  }
\label{7_to_9}
\end{figure}

To verify the latter conclusions, in 
 Fig.\ref{7_to_9}, we display plots of the Fano factor vs. $e V$,
drawn with the same values of the system parameters that we used for the plots of the scattering 
coefficients in Fig.\ref{scat_coef}.   By looking at the 
$e V \to 0$-limit of the Fano factor, we find that, the larger is $\ell$, the neater is the convergence of ${\cal P} ( 0 ) / I$ to either
0, or  $e^* = 2e$.
Importantly, the plots in Fig.\ref{7_to_9} show how it is possible to 
use a measurement of the Fano factor to detect which phase 
the LRK lies within. In the following section, we refine our analysis to spell out the behavior of 
the zero-frequency shot noise  (and, accordingly, of the Fano factor) across the  quantum critical line between
the SRTP and the LRTP, at $\alpha = 1$.

\section{The shot-noise and the Fano factor across the quantum phase transition line}
\label{crit_qua} 

\begin{figure}[ht]
\centering
\includegraphics*[width=.45\linewidth]{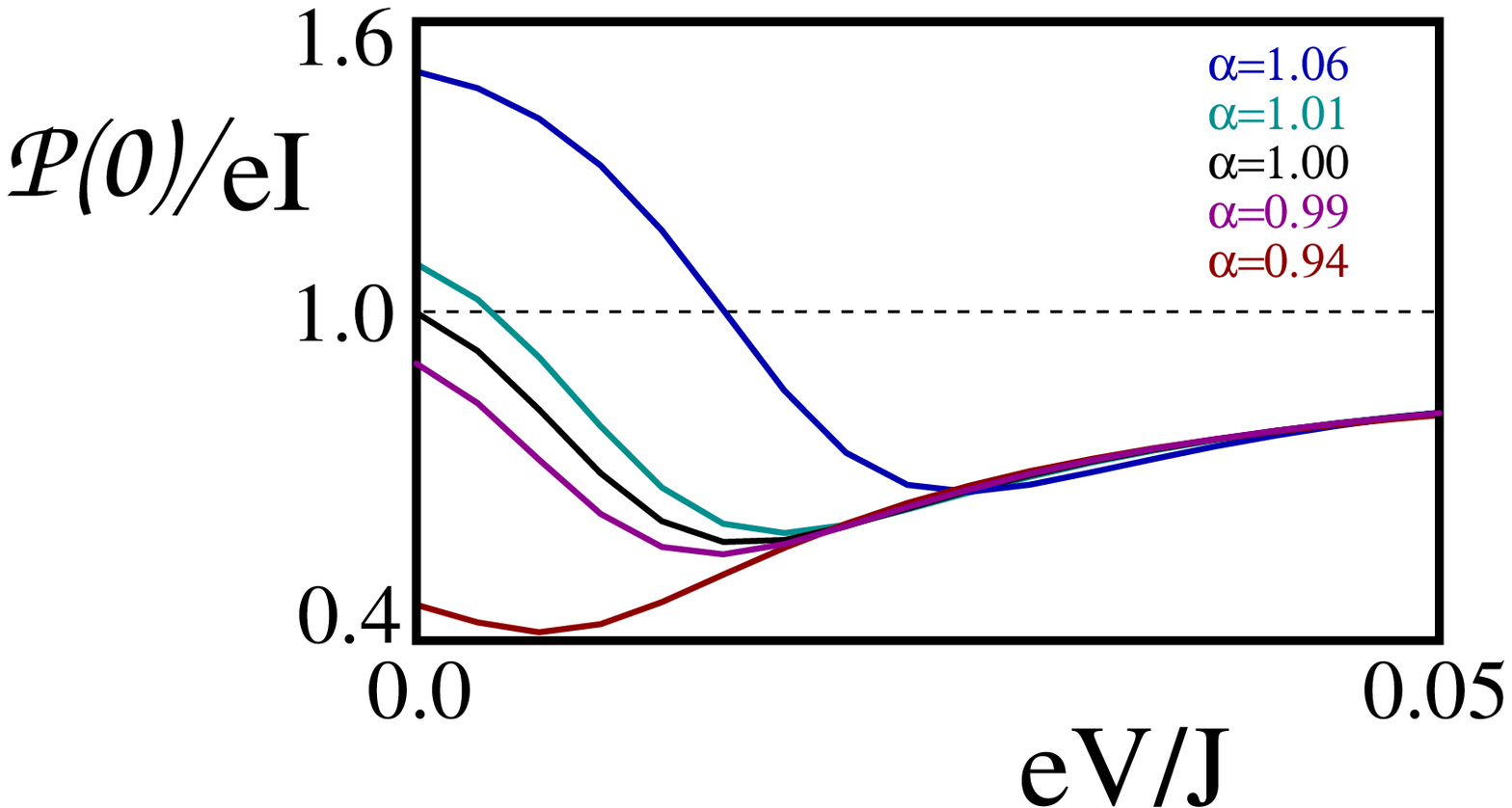}
\caption{The Fano factor  ${\cal P} ( 0  ) / I e$ computed for the central region 
Hamiltonian with   $\Delta = 2 w = 2,\mu=0.25$, $\ell = 2000$, $J=w$, and for various values of $\alpha$ close to $\alpha = 1$. The main trend clearly 
appears from the plot: At small values of $eV$ the Fano factor either converges to 0 within the SRTP, or 
it flows towards $2e$ within the LRTP.  Apparently, the more one moves from the critical value $\alpha = 1$, 
the closer the Fano factor gets to its value within 
either phase, as expected. The difficult convergence of the Fano factor for $\alpha \sim 1$ is clearly a finite-size effect.  Therefore we expect that 
the convergence can be made better by increasing $\ell$, which we avoided to do, on one hand because the trend was already 
clear enough, on the other hand because of the increasing computing time for $\ell $ larger than 2000.  An interesting 
 ''Fano factor fractionalization'' takes place as $\alpha = 1$ (see the main text for a discussion about this point).  } 
\label{critical}
\end{figure}

For the LRK,   $\alpha$ can be regarded as a sort 
of tuning parameter, by acting on which  one may in principle trigger a quantum phase transition 
between the SRTP and the LRTP.  From the previous section we expect that the Fano factor can be an efficient quantity to monitor the corresponding 
quantum phase transition (QPT). Accordingly, we now refine our analysis of the Fano factor around the critical line at   $\alpha = 1$ \cite{lepori_1,lepori_2}. 
Close to the  QPT one generically expects that, 
even within the SRTP, the overlap between the wavefunctions of the localized Majorana modes drops down really 
slowly with $\ell$ \cite{lepori_1,lepori_2}. For this reason, compared to the previous plots, we now substantially increase 
$\ell$ to $\ell = 2000$ sites. The results for the Fano factor vs. $e V$ for small values of $eV$ are reported in 
Fig.\ref{critical}.  We find that, for $\alpha > 1$ (SRTP), 
the curves bend downwards toward 0 as $e V \to 0$, consistently with the expected result that the Fano factor goes to 0, 
in the SRTP.  On the contrary, for $\alpha < 1$ (LRTP), the curves  
 bend upwards, consistently with the expected result that, within this phase,  the Fano factor tends to $2e$ as $eV \to 0$.

In order to provide a physical interpretation of the results summarized in Fig.\ref{critical}, we now make a combined use of 
the results of Ref. \cite{lepori_1} about the emergence of massive subgap modes within the LRTP (whose mass
survives the thermodynamic limit) and of the simplified model in Eq. \eqref{es.1text},  by identifying the 
parameter  $\epsilon_d$ in this specific model with  the mass of the massive subgap modes, similarly to what was done in Secs. \ref{ki_results} and \ref{ki_effectivedge}.

Within the  SRTP, the  Majorana mass goes to zero as $\ell \to \infty$. Accordingly, to address the large-$\ell$ limit for what concerns the 
transport properties of our system, we consider the scattering coefficients obtained in Appendix \ref{ssingle}, setting $\epsilon_d = 0$. 
In this case,  as $E \to 0$ we obtain 
\begin{eqnarray}
 R ( E ) &=&  \frac{E^2 }{E^2 + \Gamma^2} \nonumber \\
 A ( E ) &=&   \frac{\Gamma^2 }{E^2 + \Gamma^2} \nonumber \\
  T ( E ) &=& C ( E ) \sim 0
 \;\;\;\; , 
 \label{fsec.1}
\end{eqnarray}
\noindent
with $\Gamma = 4 t^2 \sin (k_F ) / J$.  From the explicit formula for the zero-frequency shot-noise reported in appendix \ref{transport}, 
we therefore obtain that  ${\cal P} ( 0 ) = 0$.
 The latter result  in turn implies at $T \to 0$ a noiseless current and, accordingly, a zero Fano factor, which 
explains the trend observed in Fig. \ref{critical}, as 
soon as one enters the SRTP. 

When instead $\epsilon_d$ is finite as $E \to 0$, one obtains 
(dropping the $_{L,R}$ labels, unessential because of the symmetry)
\begin{eqnarray}
r ( E ) &\approx& - \frac{\epsilon_d^2}{\epsilon_d^2 + \Gamma^2  } + {\cal O} ( E ) \, , \nonumber \\
 a ( E ) &\approx& \frac{\Gamma^2}{\epsilon_d^2 +  \Gamma^2  } + {\cal O} ( E ) \, , \nonumber \\
 t ( E ) &=& c ( E ) \approx  \frac{\epsilon_d \, \Gamma }{\epsilon_d^2 + \Gamma^2  } + {\cal O} ( E ) 
 \;\;\;\; . 
 \label{fsec.2}
\end{eqnarray}
\noindent
Using the approximate formulas in Eqs. (\ref{fsec.2}), for small values of $e V$, we obtain, as $T \to 0$,
\begin{eqnarray}
 I &=& \frac{2 e^2}{2 \pi} \; \left[ \frac{\epsilon_d^2 \, \Gamma^2}{ ( \epsilon_d^2 + \Gamma^2 )^2} \right] V \,  ,\nonumber \\
 {\cal P} ( 0  ) &=& \frac{4 e^3}{2 \pi} \; \left[ \frac{\epsilon_d^2 \, \Gamma^2}{ ( \epsilon_d^2 + \Gamma^2 )^2} \right] V 	\, , 
 \label{fsec.3}
\end{eqnarray}
\noindent
which yields the result that we inferred from the numerical calculations, that is ${\cal P} ( 0  ) / I = 2 e $
within the LRTP.  As we discuss above, this result comes from the fact that now  there is a finite rate for 
both LAR and CAR and, in addition,  a finite rate for  NR  and for
NT. In particular, in the limit of weakly coupled leads, that is $\Gamma \ll \epsilon_d < 1$, 
the dominant process is NR, the NT and the CAR  take 
place  with an intermediate weight, while the less probable process is the LAR, at both interfaces. 
Therefore, events supporting the subgap current transport (which can be regarded as due to injection/absorption of 
Cooper pairs into/from the superconducting region) become rare processes and, therefore, the corresponding fluctuations 
are expected to be Poissonian, which motivates the Fano factor becoming equal to the elementary charge 
transported through the circuit, that is $2e$. 

\begin{figure}[ht]
\centering
\includegraphics*[width=.4\linewidth]{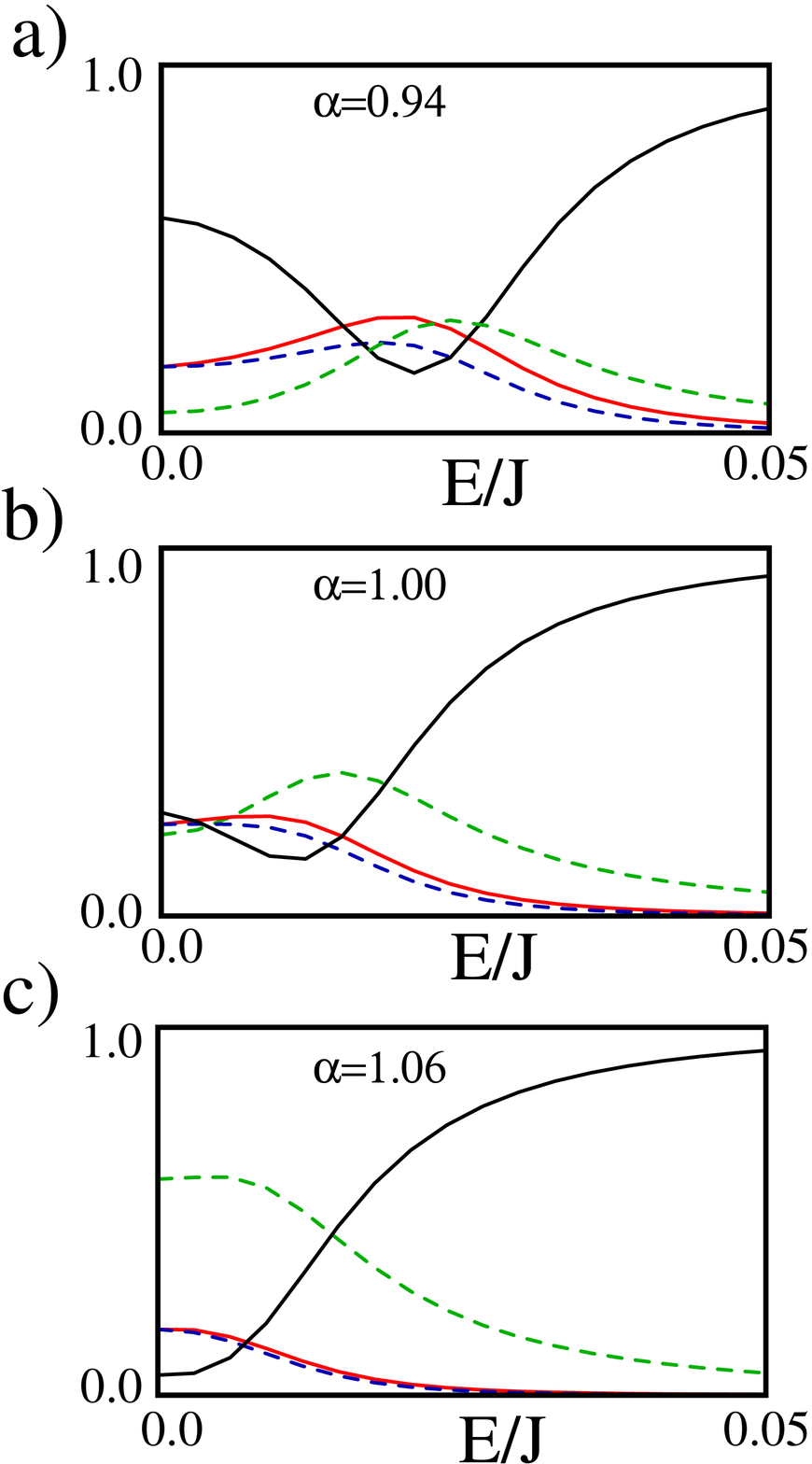}
\caption{ Scattering coefficients $ R (E) $  (black solid line), $ A(E)$ (green dashed line), 
$  T (E) $ (red solid line), and $  C ( E ) $ (blue dashed line),  computed for 
$\Delta = 2$, $ w = 1$, $\mu = 0.25$,  $k_F = \frac{\pi}{2}$, and $\ell = 2000$, at
$\alpha = 0.94$ ({\bf a)}, at $\alpha = 1.00$ ({\bf b)}, and at $\alpha = 1.06$ ({\bf c)}.}   
\label{critical2}
\end{figure}

Despite the specific example we made within the simplified model 
in Eq.\eqref{es.1text}, the discussion above applies  to more general situations, such as the LRK, provided that, as $ E \to 0$, the 
corresponding amplitudes behave consistently with the results in  Eqs.(\ref{fsec.2}).
To verify this point  for the LRK, in Fig.\ref{critical2}  we plot the scattering coefficients, computed  within the 
LRTP at $\alpha = 0.94$, at the critical value of $\alpha$, $\alpha = 1.00$ \cite{lepori_1}, and within the SRTP at 
$\alpha = 1.06$. All the plots 
have been drawn at $\ell = 2000$.  (Note that, close to the QPT, even at such value of $\ell$ the convergence of the scattering 
coefficients at   $E \to 0$ to their values in the thermodynamic limit is quite slow.  A similar slow convergence effect has been found  and discussed in 
Refs.[\onlinecite{lepori_1,lepori_5}]). 
Nevertheless,  one can already identify a well defined trend, as a function of $E$: As $\alpha = 0.94$ and $ E \to 0$,  we find that $R (E)$ takes off, which is expected, due to the 
low value of $t /J$  Eq.(\ref{eql.7})   (and of the hybridization parameter $\Gamma$).
In that limit, the lowest coefficient is  $ A ( E)$, with $ A ( E )  < C ( E ) , T ( E ) \ll R ( E ) $. 

 These results are absolutely consistent with the ones for the model in Eq.(\ref{es.1text})
(see also appendix \ref{ssingle}) at $\epsilon_d > 0$, which suggests an analogous interpretation of the low-energy 
dynamics supporting subgap current transport within the LRTP, eventually 
explaining the value ${\cal P} ( 0  ) / I = 2 e$,  found from the exact numerical calculations.  Similarly
from the plot drawn at $\alpha = 1.06$, which we report in Fig.\ref{critical}{\bf c)}, we find that 
$A ( E )$ takes off as $E \to 0$, with a corresponding suppression of all the other scattering 
coefficients. This is again consistent with the results obtained within the model in Eq. \eqref{es.1text}
as $\epsilon_d \to 0$, which implies a corresponding interpretation of the low-energy subgap dynamics, as well as 
of the numerically estimated value ${\cal P} ( 0  ) / I = 0$, within the SRTP. 

Considerably interesting per se is the plot that we show in Fig.\ref{critical}{\bf b)}, where we draw again the 
scattering coefficients computed at the critical point $\alpha = 1.00$. We find that, at this special value of 
$\alpha$, all the scattering coefficients basically converge towards an unique value as $E \to 0$. While this
numerics explains the ''Fano factor fractionalization''   at $\alpha = 1$, that is, the halving of 
the Fano factor (with respect to its value within the SRP)  to ${\cal P} (0 ) / I = e$. Such an
interesting critical fractionalization calls for a deeper investigation of the corresponding physical processes and, 
in general, of what happens  across the quantum phase transition from the  SRTP to the LRTP. Here we do not 
discuss more in detail this issue since, as it falls beyond the scope of this paper, we plan to address it in a
future publication. 

\section{Conclusions}
\label{conclusions}

In this paper we presented a possible way of monitoring the quantum phase transition between the short-range phase   and the long-range topological phase  of the 
Kitaev chain with long-range pairing, by looking at whether the mass of the emerging subgap modes flows to zero, or keeps 
finite, as the length of the chain $\ell \to \infty$. We assumed that the LRK can be contacted with two normal leads to form 
an NSN junction and we derived the low-energy behavior of the scattering coefficients across the whole junctions within 
both phases. Specifically, while within the SRTP only LAR survives at the Fermi energy, within the LRTP all the four possible scattering 
coefficients have, in general, a nonzero value at the same energy. We therefore spell out the corresponding consequences for 
the Fano factor in the small-$e V$ limit. Eventually, we prove that either the Fano factor goes to 0 as 
$e V \to 0$ within the SRTP, or it goes to $2e$ as  $e V \to 0$ within the LRTP.
We also discuss the behavior of the Fano factor at quantum critical line at $\alpha = 1$, finding the remarkable result that
it flows to $e$, as $e V \to 0$. This poses the further question of what could be the physical meaning of such a 
''fractionalization'' effect, which we plan to address in a follow-up publication.

On the practical side,  we evidence  the strict connection between the emergence of the LRTP and 
the onset of a remarkable nonzero crossed Andreev reflection at the Fermi level. On taking 
the complementary point of view in which one injects Cooper pairs into the 
circuit through the central superconducting region, this provides a potential 
source to generate nonlocal particle-hole highly entangled states, that is, thanks to the CAR our device can stabilize the emission of two correlated particles, one 
per each lead.   Therefore our model can  be regarded as an efficient generator of pairs of strongly entangled 
particles, distant in real space \cite{beenakker_1}.  Notably the described mechanism works only in the long-range regime $\alpha <1$.
 
  A separate question concerns the practical realization of our proposed   device. 
Apparently, engineering a current transport experiment requires an appropriate solid-state realization of the LRK, 
  suitable to be connected to electric leads with an applied voltage bias and allowing the measurements of  the induced electric currents and current correlations (noise).   

While a number of 
  ''optical'' experimental realizations of the LRK in appropriate atomic setups has recently emerged in the literature 
\cite{benito,rich,zol,sch_2,gop,sga,lukin2017,bettles2017}, a proposal of a solid-state realization  has been put forward in Refs.[\onlinecite{shiba_0,shiba_1}]. 
The basic idea consists in 
realizing chains of magnetic impurities on the top of a conventional s-wave superconductor and in focusing onto the dynamics 
of the subgap Shiba states emerging at the magnetic impurities themselves. In fact, Shiba states can be described as an 
effective 1D pairing Hamiltonian, with both long-range hopping, as well as long-range pairing \cite{shiba_0,shiba_1}. As
highlighted in  Refs.[\onlinecite{lepori_2,lepori_3}], adding  long-range hopping  does not give rise to qualitative modifications 
in the phase diagram, with respect to the LRK. Therefore, the effective Hamiltonian of   Refs.[\onlinecite{shiba_0,shiba_1}]  
  is expected to be qualitatively equivalent to the LRK. 
  More in detail, the long-range pairing between Shiba states 
falls with the distance $r$ as  $e^{ - r / \xi_0 }/r$, with $\xi_0$ being 
the coherence length of the host superconductor (remarkably, $\xi_0$ can be made as large as a few thousands of the lattice 
step of the chain \cite{shiba_0,shiba_1}). Therefore, in chains of length $\ell \gg \xi_0$ this is expected to 
mimic pairing in the SRK, while chains of length $\ell \ll \xi_0$ are expected to behave as a LRK with $\alpha = 1$. 
As discussed above, $\alpha = 1$ corresponds to the critical regime, where the Fano factor should jump from 0 to a finite value
$\sim e$. Accordingly, as $\xi_0$  is in principle a tunable parameter by acting, for instance, with an applied magnetic field, or by 
changing the temperature,  it can be used to drive the system (at fixed $\ell$)  between an effectively short-range pairing regime 
and a critical regime at $\alpha = 1$, the latter one being already characterized by 
long-range correlations (witnessed by the finite value of the Fano factor as $V \to 0$). Note that keeping 
$\ell$ finite does not constitute an obstruction for the emergence of long-range physics, as described in detail in Ref.[\onlinecite{bradipo}].
While this system does not appear to rigorously map 
onto the LRK at $\alpha \geq 1$, it is likely that, for what concerns the phase diagram, acting on $\xi_0$ in a finite-length $\ell$ chain should be qualitatively equivalent
to vary $\alpha$ in the LRK. Finally, by tunnel-coupling the endpoints of the chain to conducting quantum wires, one  is expected to readily realize the whole circuit 
we propose to use to measure the current and the shot-noise across the LRK.
  
Of course, it is still unclear how to realize in a solid state system the long-range pairing phase at  $\alpha < 1$ strictly. 
However, we are confident that   the enormous and continuous progress in the engineering of nanostructures 
and nanodevices is likely to soon make it available or, alternatively, to enable experimentalists to design  appropriate ''optical'' 
versions of the experiment in pertinently designed atomic setups \cite{benito,rich,zol,sch_2,gop,sga,lukin2017,bettles2017} .

\vspace{0.5cm}

{\bf Acknowledgements --} 
We thank M. Burrello, G. Campagnano, A. Nava, A. Tagliacozzo, and A. Trombettoni for useful discussions. 
S. P. acknowledges support by a Rita Levi-Montalcini fellowship of the Italian MIUR.

\appendix 

\section{Construction of the fully dressed imaginary-time Green's function}
\label{construction}

In this section we review the derivation of the exact Green's function in the  
real space-imaginary frequency representation, ${\bf C}_{ j , j'} ( i \omega )$ \cite{mahan}. For the sake of
discussion, we add to the ${\bf C}$ function an additional pair of labels, $X , X' = L , R$, which 
evidence whether $j $ and $j'$ refer to sites within the left-hand, or the right-hand normal lead. Using 
the same notation of the main text, we therefore set 
\beq
{\bf C}_{ ( X, X' ) ; (j , j')} ( \tau ) = - \left[ \begin{array}{cc}
\langle  {\bf T}_\tau [ c_{ X , j} ( \tau ) c_{ X' , j'}^\dagger ( 0 ) ] \rangle & 
\langle  {\bf T}_\tau [ c_{ X , j}^\dagger ( \tau ) c_{ X' , j'}^\dagger ( 0 ) ] \rangle \\
\langle  {\bf T}_\tau [ c_{ X , j}( \tau ) c_{ X' , j'} ( 0 ) ] \rangle &
\langle  {\bf T}_\tau [ c_{ X , j}^\dagger  ( \tau ) c_{ X' , j'} ( 0 ) ] \rangle
                                                    \end{array} \right]
\:\:\:\: , 
\label{ap1.1}
\eneq
\noindent
and, accordingly 
\beq
{\bf C}_{ ( X , X' ) ; ( j , j' ) } ( i \omega ) = \int_0^\beta \: d \tau \: e^{ i \omega \tau} 
\: {\bf C}_{ ( X, X' ) ; (j , j')} ( \tau )
\:\:\:\: . 
\label{ap1.2}
\eneq
\noindent

\subsection{The Green's function for the disconnected leads}
\label{gfdisco}

To regularize the calculations, we consider leads made of 
$\Lambda$ sites, eventually sending $\Lambda \to \infty$. Accordingly, the left-hand lead consists of 
sites running from $j = - \Lambda + 1$ to $j = 0$, while the right-hand lead consists of 
sites running from $j=\ell + 1$ to $j = \ell +  \Lambda$. Accordingly,  write the 
lead Hamiltonians as 
\beq
{\cal H}_{R , L } = \sum_k \: \xi_k \: c_{ (L , R ) , k}^\dagger c_{ ( L , R ) , k }
\;\;\;\; ,
\label{supercond.7}
\eneq
\noindent
with $\xi_k = - 2 J \cos ( k ) - \mu$ and $k = \frac{\pi n }{\Lambda + 1}$, with $n = 1 , \ldots , \Lambda$, and 
\begin{eqnarray}
 c_{ L , k } &=& \sqrt{\frac{2}{\Lambda + 1}} \: \sum_{ j = - \Lambda + 1}^0  \: \sin [ k ( j + \Lambda )  ] \: c_{ L , j }   
 \equiv \sum_{ j = - \Lambda + 1}^0  \: \phi_k ( j + \Lambda ) \: c_{ L , j } \, \, , \nonumber \\
 c_{ R , k } &=& \sqrt{\frac{2}{\Lambda + 1}} \: \sum_{ j = \ell + 1}^{\Lambda + \ell  }   \: \sin [ k ( j - \ell  )  ] \: c_{ R , j }
 \equiv   \: \sum_{ j = \ell + 1}^{\Lambda + \ell }  \phi_k ( j - \ell )  c_{ R , j }\;  \:\:\:\: .
 \label{supercond.8}
\end{eqnarray}
\noindent
It is, now, straightforward, though tedious, to compute the Green's functions 
${\bf C}_{(L,L);(j,j')}^{(0)} ( i \omega )$ and ${\bf C}_{(R ,R );(j,j')}^{(0)} ( i \omega )$.
The result is 
\begin{eqnarray}
{\bf C}_{ (L , L ) ; (j , j' )  }^{(0)} ( i \omega ) &=& \sum_k \phi_{L , k} ( j ) \phi_{L , k}( j' ) \: 
\left[ \begin{array}{cc}
( i \omega - \xi_k )^{-1} & 0 \\ 0 & ( i \omega + \xi_k )^{-1}                                                         
                                                       \end{array} \right] = \left[�\begin{array}{cc}  G_{ ( L , L )  ; (j , j') }^{(0)}  ( i \omega )  & 0 \\ 
0 &  - [G_{ ( L , L )  ; (j , j') }^{(0)}  (   i \omega ) ]^*
\end{array} \right]
\nonumber \\
{\bf C}_{ (R , R ) ; (j , j' )  }^{(0)}( i \omega ) &=& \sum_k \phi_{R , k} ( j ) \phi_{R , k}( j' ) \: 
\left[ \begin{array}{cc}
( i \omega - \xi_k )^{-1} & 0 \\ 0 & ( i \omega + \xi_k )^{-1}                                                         
                                                       \end{array} \right] = \left[�\begin{array}{cc}  G_{ ( R , R )  ; (j , j') }^{(0)}  ( i \omega )  & 0 \\ 
0 &  - [ G_{ ( R , R )  ; (j , j') }^{(0)}  (   i \omega ) ]^*
\end{array} \right]
 \:\:\:\: ,
 \label{gsuper.2}
 \end{eqnarray}
 \noindent
with 
\begin{eqnarray}
G_{ ( L , L ) ; ( j , j' )}^{(0)} ( i \omega ) &=& \frac{1}{2 J } \: \left\{ \frac{\sinh [\lambda ( \omega ) ( \Lambda + 1 - | j - j' | ) ]-
\sinh [\lambda ( \omega ) ( \Lambda - 1 + j + j ' ) ] }{ \sinh [ \lambda ( \omega ) ( \Lambda +1 ) ] \sinh [ \lambda ( \omega ) ] } \right\}
\nonumber \\
G_{ ( R  , R  ) ; ( j , j' )}^{(0)} ( i \omega ) &=& \frac{1}{2 J } \: \left\{ \frac{\sinh [\lambda ( \omega ) ( \Lambda + 1 - | j - j' | ) ]-
\sinh [\lambda ( \omega ) ( \Lambda + 1 + 2 \ell -  j -  j ' ) ] }{ \sinh [ \lambda ( \omega ) ( \Lambda +1 ) ] \sinh [ \lambda ( \omega ) ] } \right\}
\:\:\:\: ,
\label{me.3}
\end{eqnarray}
\noindent
and $e^{\pm \lambda ( \omega )}$ being the roots of the algebraic equation  $z^2 +  2 \left( \frac{\mu+ i \omega}{2 J } \right) z + 1 = 0 $.
If only low values of $\omega$ are relevant to the calculations, then $\lambda ( \omega )$ can be 
expanded according to $ \lambda ( \omega ) \approx i k_F - \frac{ \omega }{v}$.
As a result, we may approximate Eqs.(\ref{me.3}) as 
\begin{eqnarray}
G_{ ( L , L ) ; ( j , j' )}^{(0)} ( i \omega ) &\approx & \frac{1}{2 J } \: \left\{ \frac{ \sinh 
\left[ \left( i k_F - \frac{ \omega}{v} \right) ( \Lambda + 1 - | j - j' | ) \right] -
\sinh \left[ \left( i k_F - \frac{ \omega}{v} \right)  
( \Lambda - 1 + j + j ' ) \right] }{ \sinh \left[ \left( i k_F - \frac{ \omega}{v} \right) ( \Lambda +1 ) \right] 
\sinh \left[ \left( i k_F - \frac{ \omega}{v} \right)  \right] } \right\}
\nonumber \\
G_{ ( R  , R  ) ; ( j , j' )}^{(0)} ( i \omega ) &\approx& \frac{1}{2 J } \: \left\{ \frac{ \sinh 
\left[ \left( i k_F - \frac{ \omega}{v} \right) ( \Lambda + 1 - | j - j' | ) \right] -
\sinh \left[ \left( i k_F - \frac{ \omega}{v} \right)  ( \Lambda + 1 + 2 \ell -  j -  j ' ) \right] }{ 
\sinh \left[ \left( i k_F - \frac{ \omega}{v} \right) ( \Lambda +1 ) \right] 
\sinh \left[ \left( i k_F - \frac{ \omega}{v} \right)  \right] } \right\}
\:\:\:\: . 
\label{me.5}
\end{eqnarray}
\noindent
In the following, we use the quantities we computed in Eqs.(\ref{me.3},\ref{me.5}) as building blocks to 
construct the fully dressed Green's functions in the mixed representation.

\subsection{Fully dressed Green's functions in the mixed representation}
\label{fully_dress}

In computing the exact Green's functions of the system, a necessary ingredient is the 
Green's functions for the central region S, ${\bf G}_{d ; (r,r')} ( \tau )$. In 
Nambu representation, this is given by 
\beq
{\bf G}_{ d ; ( r, r') } ( \tau ) = - \left[ \begin{array}{cc}
\langle {\bf T}_\tau [ d_r ( \tau ) d_{r'}^\dagger ( 0 ) ] \rangle &  \langle {\bf T}_\tau [ d_r^\dagger ( \tau ) d_{r'}^\dagger ( 0 ) ] \rangle       \\
\langle {\bf T}_\tau [ d_r  ( \tau ) d_{r'}  ( 0 ) ] \rangle  &  \langle {\bf T}_\tau [ d_r^\dagger ( \tau ) d_{r'} ( 0 ) ] \rangle  
                                             \end{array} \right]
\;\;\;\; , 
\label{me.6}
\eneq
\noindent
and, as usual, we define its Fourier transform as 

\beq
{\bf G}_{ d ; (r,r')} ( i \omega ) = \int_0^\beta \: d \tau \: e^{ i \omega \tau} \: {\bf G}_{ d ; ( r, r') } ( \tau ) 
\:\:\:\: . 
\label{me.7}
\eneq
\noindent
In terms of the solutions of the Bogoliubov-de Gennes equations within the central region with open boundary 
conditions, $( u_r , v_r)$, with $r = 1 , \ldots , \ell$,   one may represent ${\bf G}_{ d ; (r,r')} ( i \omega ) $ as
\beq
{\bf G}_{ d ; (r,r')} ( i \omega ) = \sum_{ \epsilon > 0 } \left\{ \frac{1}{i \omega - \epsilon} \: 
\left[ \begin{array}{cc}
u^\epsilon_r (u^\epsilon_{r'} )^* &   u^\epsilon_r (v^\epsilon_{r'} )^*    \\  v^\epsilon_r (u^\epsilon_{r'} )^* &  v^\epsilon_r (v^\epsilon_{r'} )^* 
       \end{array} \right] +  \frac{1}{i \omega + \epsilon} \: 
\left[ \begin{array}{cc}
( v^\epsilon_r )^* v^\epsilon_{r'}  &   ( v^\epsilon_r )^* u^\epsilon_{r'}    \\  ( u^\epsilon_r )^* v^\epsilon_{r'}  &   (u^\epsilon_r )^* u^\epsilon_{r'}  
       \end{array} \right] \right\}
\;\;\;\; . 
\label{meme.1}
\eneq
\noindent
By means of a systematic application of the equation of motion approach, one may readily derive the 
full set of Dyson's equations for the fully dressed Green's functions ${\bf C}_{ ( X , X' ) ; ( j , j' ) } ( i \omega )$. 
These are given by  
\begin{eqnarray}
 {\bf C}_{(L , L ) ; ( j , j')} ( i \omega) &=&  {\bf C}^{(0)}_{(L , L ) ; ( j , j')} ( i \omega) + t^2 {\bf C}_{ ( L , L ); (j , 0 )}^{(0)} ( i \omega) 
 \tau^z  {\bf G}_{ d ; (1,1)} ( i \omega ) \tau^z  {\bf C}_{(L , L ) ; ( 0 , j')} ( i \omega) \nonumber \\
 &+& t^2 {\bf C}_{ ( L , L ); (j , 0 )}^{(0)} ( i \omega) 
 \tau^z  {\bf G}_{ d ; (1, \ell )} ( i \omega ) \tau^z {\bf C}_{(R , L ) ; ( \ell + 1 , j')} ( i \omega) \nonumber \\
  {\bf C}_{(L , R ) ; ( j , j')} ( i \omega) &=&    t^2 {\bf C}_{ ( L , L ); (j , 0 )}^{(0)} ( i \omega) 
 \tau^z {\bf G}_{ d ; (1,1)} ( i \omega )  \tau^z {\bf C}_{(L , R ) ; ( 0 , j')} ( i \omega) \nonumber \\
 &+& t^2 {\bf C}_{ ( L , L ); (j , 0 )}^{(0)} ( i \omega) 
 \tau^z {\bf G}_{ d ; (1, \ell )} ( i \omega )  \tau^z {\bf C}_{(R , R ) ; ( \ell + 1 , j')} ( i \omega) \nonumber \\
  {\bf C}_{(R , L ) ; ( j , j')} ( i \omega) &=&  t^2 {\bf C}_{ ( R , R ); (j , \ell + 1 )}^{(0)} ( i \omega) 
 \tau^z {\bf G}_{ d ; (\ell ,1)} ( i \omega )  \tau^z  {\bf C}_{(L , L ) ; ( 0  , j')} ( i \omega) \nonumber \\
 &+& t^2 {\bf C}_{ ( R , R ); (j , \ell + 1 )}^{(0)} ( i \omega) 
 \tau^z {\bf G}_{ d ; (\ell , \ell )} ( i \omega )  \tau^z  {\bf C}_{(R , L ) ; ( \ell + 1 , j')} ( i \omega) \nonumber \\
  {\bf C}_{(R , R ) ; ( j , j')} ( i \omega) &=&  {\bf C}^{(0)}_{(R , R ) ; ( j , j')} ( i \omega) + t^2 {\bf C}_{ ( R , R ); (j , \ell + 1 )}^{(0)} ( i \omega) 
 \tau^z {\bf G}_{ d ; (\ell ,1)} ( i \omega )  \tau^z {\bf C}_{(L , R ) ; ( 0 , j')} ( i \omega) \nonumber \\
 &+& t^2 {\bf C}_{ ( R , R ); (j , \ell + 1  )}^{(0)} ( i \omega) 
 \tau^z {\bf G}_{ d ; (\ell , \ell )} ( i \omega )  \tau^z  {\bf C}_{(R , R ) ; ( \ell + 1 , j')} ( i \omega)
 \:\:\:\: ,
 \label{supercond.21}
\end{eqnarray}
\noindent
with $\tau^z$ being a Pauli matrix acting within the Nambu space. 
Taking into account that one gets ${\bf C}_{(L,L);(0,0)}^{(0)} ( i \omega ) = {\bf C}_{(R , R);(\ell + 1 ,\ell + 1)}^{(0)} ( i \omega )$,
one may formally write down 
\begin{eqnarray}
&& \left[ \begin{array}{c} 
        {\bf C}_{(L,L);(0,j')} ( i \omega ) \\ 
            {\bf C}_{(R,L);(\ell + 1 ,j')} ( i \omega ) 
       \end{array} \right] = 
 \left[ \begin{array}{c} 
        {\bf C}_{(L,L);(0,j')}^{(0)}  ( i \omega ) \\ 
            {\bf 0} 
       \end{array} \right]   
       + \nonumber \\
&& t^2 \left[ \begin{array}{cc}
   {\bf C}_{(L,L);(0,0)}^{(0)}  ( i \omega )   & {\bf 0} \\
 {\bf 0}   &   {\bf C}_{(L,L);(0,0)}^{(0)}  ( i \omega ) 
           \end{array} \right] 
\left[ \begin{array}{cc}
\tau^z {\bf G}_{d;(1,1)} ( i \omega ) \tau^z &     \tau^z {\bf G}_{d;(1,\ell )} ( i \omega ) \tau^z     \\
\tau^z {\bf G}_{d;(\ell ,1)} ( i \omega ) \tau^z &     \tau^z {\bf G}_{d;(\ell ,\ell )} ( i \omega ) \tau^z  
       \end{array} \right] 
   \left[ \begin{array}{c} 
        {\bf C}_{(L,L);(0,j')} ( i \omega ) \\ 
            {\bf C}_{(R,L);(\ell + 1 ,j')} ( i \omega ) 
       \end{array} \right]
\:\:\:\: ,
\label{sssc.1}
\end{eqnarray}
\noindent
and 
\begin{eqnarray}
&& \left[ \begin{array}{c} 
        {\bf C}_{(L,R);(0,j')} ( i \omega ) \\ 
            {\bf C}_{(R,R);(\ell + 1 ,j')} ( i \omega ) 
       \end{array} \right] = 
 \left[ \begin{array}{c} {\bf 0} \\  
        {\bf C}_{(R,R);(\ell + 1 ,j')}^{(0)}  ( i \omega ) 
       \end{array} \right]   
       + \nonumber \\
&& t^2 \left[ \begin{array}{cc}
   {\bf C}_{(L,L);(0,0)}^{(0)}  ( i \omega )   & {\bf 0} \\
 {\bf 0}   &   {\bf C}_{(L,L);(0,0)}^{(0)}  ( i \omega ) 
           \end{array} \right] 
\left[ \begin{array}{cc}
\tau^z {\bf G}_{d;(1,1)} ( i \omega ) \tau^z &     \tau^z {\bf G}_{d;(1,\ell )} ( i \omega ) \tau^z     \\
\tau^z {\bf G}_{d;(\ell ,1)} ( i \omega ) \tau^z &     \tau^z {\bf G}_{d;(\ell ,\ell )} ( i \omega ) \tau^z  
       \end{array} \right] 
   \left[ \begin{array}{c} 
        {\bf C}_{(L,R);(0,j')} ( i \omega ) \\ 
            {\bf C}_{(R,R);(\ell + 1 ,j')} ( i \omega ) 
       \end{array} \right]
\:\:\:\: .
\label{sssc.2}
\end{eqnarray}
\noindent
We now define the matrix ${\bf M}  ( i \omega )$ as 
\begin{eqnarray}
 && {\bf M}  ( i \omega ) = \left\{ {\bf I} - t^2 \left[ \begin{array}{cc}
   {\bf C}_{(L,L);(0,0)}^{(0)}  ( i \omega )   & {\bf 0} \\
 {\bf 0}   &   {\bf C}_{(L,L);(0,0)}^{(0)}  ( i \omega ) 
           \end{array} \right] 
\left[ \begin{array}{cc}
\tau^z {\bf G}_{d;(1,1)} ( i \omega ) \tau^z &     \tau^z {\bf G}_{d;(1,\ell )} ( i \omega ) \tau^z     \\
\tau^z {\bf G}_{d;(\ell ,1)} ( i \omega ) \tau^z &     \tau^z {\bf G}_{d;(\ell ,\ell )} ( i \omega ) \tau^z  
       \end{array} \right]  \right\}^{-1} \nonumber \\
       && \equiv 
       \left[ \begin{array}{cc}
M_{1,1} ( i \omega ) & M_{1,2 } ( i \omega )\\ 
M_{2, 1} ( i \omega )  & M_{2,2 } ( i \omega )
              \end{array}
\right]
\:\:\:\: . 
\label{sssc.3}
\end{eqnarray}
\noindent
As a result, we obtain 
\begin{eqnarray}
 {\bf C}_{ ( L,L) ; ( j , j' ) } ( i \omega ) &=&  {\bf C}_{ ( L,L) ; ( j , j' ) }^{(0)}  ( i \omega ) + 
 {\bf C}_{ ( L,L) ; ( j ,0  ) }^{(0)}  ( i \omega ) \Upsilon_{(L,L)} ( i \omega )   {\bf C}_{ ( L,L) ; ( 0 , j' ) }^{(0)} ( i \omega )
 \nonumber \\
       {\bf C}_{(R , L ) ; ( j  , j')} ( i \omega) &=&  {\bf C}_{(R,R) ; ( j , \ell + 1   ) }^{(0)} ( i \omega )
        \Upsilon_{(R,L)} ( i \omega )   {\bf C}_{ ( L,L) ; ( 0 , j' ) }^{(0)}  ( i \omega )\nonumber \\
 {\bf C}_{(L  , R ) ; ( j  , j')} ( i \omega) &=&    {\bf C}_{ ( L,L) ; ( j , 0 ) }^{(0)}   ( i \omega )   \Upsilon_{( L , R)} ( i \omega )   
 {\bf C}_{ ( R,R) ; ( \ell + 1 , j' ) }^{(0)}  ( i \omega )   \nonumber \\
  {\bf C}_{ ( R,R) ; ( j , j' ) } ( i \omega ) &=&  {\bf C}_{ ( R,R) ; ( j , j' ) }^{(0)}  ( i \omega )+ 
  {\bf C}_{ (R, R ) ; ( j , \ell + 1   ) }^{(0)}  ( i \omega ) \Upsilon_{(R , R)} ( i \omega )   {\bf C}_{ ( R, R) ; ( \ell + 1 , j' ) }^{(0)} ( i \omega )
 \:\:\:\: , 
 \label{ultimate.2}
\end{eqnarray}
\noindent
with 
\begin{eqnarray}
  \Upsilon_{(L,L)} ( i \omega )  &=&   t^2  \tau^z   {\bf G}_{ d ; (1,1)} ( i \omega ) \tau^z M_{1,1} ( i \omega ) +  
  t^2   \tau^z  {\bf G}_{ d ; (1,\ell)} ( i \omega ) \tau^z  M_{2,1}  ( i \omega )  \nonumber \\
  \Upsilon_{(R,L)} ( i \omega )    &=&   t^2   \tau^z  {\bf G}_{ d ; (\ell ,1 )} ( i \omega ) \tau^z M_{1,1}  ( i \omega )  + 
 t^2    \tau^z {\bf G}_{ d ; (\ell , \ell )} ( i \omega )\tau^z M_{2,1}  ( i \omega )  \nonumber \\
  \Upsilon_{( L , R)} ( i \omega )  &=&  t^2   \tau^z  {\bf G}_{ d ; (1,1)}  ( i\omega)  \tau^z M_{1,2}  ( i \omega )  + 
 t^2    \tau^z  {\bf G}_{ d ; (1,\ell )} ( i \omega )\tau^z   M_{2,2}  ( i \omega )  \nonumber \\
  \Upsilon_{(R , R)} ( i \omega )   &=&  t^2   \tau^z  {\bf G}_{ d ; (\ell ,1)} ( i \omega ) \tau^z M_{1,2}  ( i \omega )  + 
   t^2   \tau^z  {\bf G}_{ d ; (\ell ,\ell)} ( i \omega )\tau^z M_{2,2}  ( i \omega )  \, \, .
 \label{ultimate.3}
 \end{eqnarray}
\noindent
Equations (\ref{ultimate.3}) encode  the Green's functions of the leads for the fully dressed 
system. To present the fully dressed Green's functions in a form explicitly containing the scattering amplitudes, we 
have still to send $\Lambda \to \infty$, which is the subject of the next subsection.

\subsection{Fully dressed Green's functions in terms of the scattering amplitudes}
\label{grescat}

Before taking the $\Lambda \to \infty$ limit, we recall that, for the sake of performing the analytical continuation to 
real frequencies, we have to work at $\omega > 0$. Making such an assumption, we send $\Lambda \to \infty$ in 
the Green's functions we derived before, obtaining  
 
\begin{eqnarray}
{\bf C}_{(L,L); (j,j')} ( i \omega ) &=& \left[\begin{array}{cc} \frac{ e^{   \left( - i k_F + \frac{\omega}{v}\right) | j - j' | }  }{v ( i \omega ) } & 0 \\
0 & \frac{ e^{  \left( - i k_F - \frac{\omega}{v}�\right) | j - j' | }  }{v ( - i \omega ) }  \end{array} \right]\nonumber \\
&+& e^{ \frac{\omega}{v}( j + j' - 2 ) } \: \left[�\begin{array}{cc}e^{- i k_F ( j + j' - 2 ) } \left[�- \frac{1}{v ( i \omega ) }+ \frac{\Upsilon_{L,L}^{(1,1)} 
( i \omega ) }{J^2} \right] & - e^{ -i k_F ( j - j'  ) }   \frac{\Upsilon_{L,L}^{(1,2)} 
( i \omega ) }{J^2} \\
 - e^{  i k_F ( j - j'  ) }   \frac{\Upsilon_{L,L}^{(2,1)} 
( i \omega ) }{J^2}  & e^{   i k_F ( j + j' - 2 ) } \left[- \frac{1}{v ( - i \omega ) }+ \frac{\Upsilon_{L,L}^{(2,2)} 
( i \omega ) }{J^2} \right]  \end{array}\right] \, \, \, ,
\:\:\:\: , 
\label{grs.6}
\end{eqnarray}
\noindent 
\beq
{\bf C}_{(L,R); (j,j')} ( i \omega ) =  e^{ \frac{\omega}{v}( j - j' + \ell -1  ) } \: 
\left[\begin{array}{cc}e^{ - i k_F ( j - j' + \ell -1) }   \frac{\Upsilon_{L,R}^{(1,1)} 
( i \omega ) }{J^2}  & - e^{ - i k_F ( j+ j' - \ell - 1  ) }   \frac{\Upsilon_{L,R}^{(1,2)} 
( i \omega ) }{J^2} \\
 - e^{ i k_F ( j +  j' - \ell - 1  ) }   \frac{\Upsilon_{L, R }^{(2,1)} 
( i \omega ) }{J^2}  & e^{   i k_F ( j  -  j'  + \ell -1 ) }  \frac{\Upsilon_{L,R}^{(2,2)} 
( i \omega ) }{J^2}   \end{array}\right] 
 \;\;\;\; , 
 \label{grs.7}
 \eneq
 \noindent
\beq
{\bf C}_{(R,L); (j,j')} ( i \omega ) =  e^{ - \frac{\omega}{v}( j - j' -  \ell  +1 ) } \: 
\left[\begin{array}{cc}e^{  i k_F ( j - j' -  \ell +1 ) }   \frac{\Upsilon_{R,L}^{(1,1)} 
( i \omega ) }{J^2}  & - e^{   i k_F ( j+ j' - \ell - 1  ) }   \frac{\Upsilon_{R,L}^{(1,2)} 
( i \omega ) }{J^2} \\
 - e^{  -  i k_F ( j +  j' - \ell - 1  ) }   \frac{\Upsilon_{R, L }^{(2,1)} 
( i \omega ) }{J^2}  & e^{ -  i k_F ( j  -  j'  - \ell +1   ) }  \frac{\Upsilon_{R,L}^{(2,2)} 
( i \omega ) }{J^2}   \end{array}\right]
 \;\;\;\; , 
 \label{grs.8}
 \eneq
 \noindent
 and, finally 
  \begin{eqnarray}
{\bf C}_{(R , R); (j,j')} ( i \omega ) &=& \left[\begin{array}{cc} \frac{ e^{ - \left( - i k_F + \frac{\omega}{v}\right) | j - j' | }  }{v ( i \omega ) } & 0 \\
0 & \frac{ e^{  \left(- i k_F - \frac{\omega}{v}\right) | j - j' | }  }{v ( - i \omega ) }  \end{array} \right]\nonumber \\
&+& e^{-  \frac{\omega}{v}( j + j' - 2 \ell ) } \: \left[\begin{array}{cc}e^{ - i k_F ( j + j' - 2\ell ) } \left[- \frac{1}{v ( i \omega ) }+ \frac{\Upsilon_{R,R}^{(1,1)} 
( i \omega ) }{J^2} \right] & - e^{  i k_F ( j - j'  ) }   \frac{\Upsilon_{R,R}^{(1,2)} 
( i \omega ) }{J^2} \\
 - e^{  - i k_F ( j - j'  ) }   \frac{\Upsilon_{R , R}^{(2,1)} 
( i \omega ) }{J^2}  & e^{   i k_F ( j + j' - 2 \ell  ) } \left[- \frac{1}{v ( - i \omega ) }+ \frac{\Upsilon_{R,R}^{(2,2)} 
( i \omega ) }{J^2} \right]  \end{array}\right] \, \, \, . 
\label{grs.9}
\end{eqnarray}
\noindent 
Eqsuations (\ref{grs.6}), (\ref{grs.7}), (\ref{grs.8}), and (\ref{grs.9}) provide us with the necessary means to 
extract the imaginary-time scattering amplitudes from the Green's function, which is the subject of 
the next appendix. 

\section{Relations between the Green's function and the scattering amplitudes across the central region}
\label{scattering_green}

In this appendix, we review the derivation of the relation between the fully dressed Green's functions of the 
NSN junction and the scattering amplitudes across the central region. At a given energy,  the four independent solutions of 
the Bogoliubov-de Gennes equations
for the whole NSN junction satisfying the proper scattering boundary conditions are:  

\subsection{Solutions obeying scattering boundary conditions} 
\label{scaboucon}
 
Referring to the asymptotic form of the solutions of the BdG equations at energy $E$ that satisfy the scattering 
boundary conditions, that is  for either $j$ belonging to the left-hand lead ($j\leq 0$) or 
for $j$ belonging to the right-hand lead ($j \geq \ell + 1$), there are four 
possible solutions, which we label in the following according to  whether the incoming state is a particle-like or a hole-like state and to whether 
the particle/hole comes from the left-hand or from the right-hand lead. We therefore consider the following solutions:
\begin{itemize}
 \item {\bf Incoming particle from the left-hand lead}
 
The corresponding eigenfunction is given by  
  \beq
 \left[ \begin{array}{c}
         u_j^E  \\ v_j^E  
        \end{array} \right]_{(p,+)} =  c_{p , + } \: \left[ \begin{array}{c}
                                                             e^{ i k_p ( E ) (j -1) } + r_{L,L} ( E ) e^{ - i k_p ( E ) ( j -1) } \\
                                                             a_{ L,L} ( E ) e^{ i k_h ( E ) (j - 1) } 
                                                            \end{array} \right]
                                                            \;\;\; , \;\; ({\rm for} \: j \leq 0 )
\;\;\;\; , 
\label{addiz.1}
\eneq
\noindent
and 
\beq
 \left[ \begin{array}{c}
         u_j^E  \\ v_j^E  
        \end{array} \right]_{(p,+)} =  c_{p , + } \: \left[ \begin{array}{c}
                                                              t_{R,L} ( E ) e^{   i k_p ( E ) ( j - \ell )  } \\
                                                             c_{ R,L} ( E ) e^{ - i k_h ( E ) (j - \ell )  } 
                                                            \end{array} \right]
                                                            \;\;\; , \;\; ({\rm for} \: j \geq \ell + 1 )
\;\;\;\; . 
\label{addiz.2}
\eneq
\noindent
Referring to the model Hamiltonians in Eqs.(\ref{eql.1}) for the leads, we define the particle momentum 
at energy $E$, $k_p ( E )$, and the hole momentum at the same energy, $k_h ( E )$, as 
\beq
E = - 2 J \cos [ k_p ( E ) ] - \mu = \mu + 2 J \cos [ k_h ( E ) ]
\;\;\;\; .
\label{addiz.1a}
\eneq
\noindent
Finally,  $c_{p , +}$ (as well as $c_{p,-}, c_{h,+}$, and $c_{h, -}$ below)
 denotes an appropriate  overall normalization constant.

 \item {\bf Incoming particle from the right-hand lead}

The corresponding eigenfunction is given by  
 \beq
 \left[ \begin{array}{c}
         u_j^E  \\ v_j^E  
        \end{array} \right]_{(p,-)} =  c_{p , -  } \: \left[ \begin{array}{c}
                                                              t_{L,R} ( E ) e^{   - i k_p ( E ) (j - 1) } \\
                                                             c_{ L,R} ( E ) e^{ i k_h ( E ) (j - 1) } 
                                                             \end{array} \right] \;\;\; , \;\; ({\rm for} \: j \leq 0 )
\:\:\:\: , 
\label{addiz.3}
\eneq
\noindent
and 
\beq
 \left[ \begin{array}{c}
         u_j^E  \\ v_j^E  
        \end{array} \right]_{(p,-)} =  c_{p , - } \: \left[ \begin{array}{c}
                                                             e^{ -  i k_p ( E ) (  j - \ell )  } + r_{R,R} ( E ) e^{  i k_p ( E ) (j - \ell )  } \\
                                                             a_{ R,R} ( E ) e^{ - i k_h ( E ) ( j - \ell )  } 
                                                            \end{array} \right]
                                                            \;\;\; , \;\; ({\rm for} \: j \geq \ell + 1 )
\;\;\;\; ; 
\label{addiz.4}
\eneq
\noindent
 
 \item {\bf Incoming hole from the left-hand lead}

 The corresponding eigenfunction is given by  
  \beq
 \left[ \begin{array}{c}
         u_j^E  \\ v_j^E  
        \end{array} \right]_{(h,+)} =  c_{h , + } \: \left[ \begin{array}{c}
                                                             \tilde{a}_{L,L} ( E ) e^{ - i k_p ( E ) ( j - 1)  } \\
                                                           e^{-  i k_h ( E ) ( j - 1)  } +   \tilde{r}_{ L,L} ( E ) e^{ i k_h ( E ) (  j - 1)  } 
                                                            \end{array} \right]
                                                            \;\;\; , \;\; ({\rm for} \: j \leq 0)
\;\;\;\; , 
\label{addiz.5}
\eneq
\noindent
and 
\beq
 \left[ \begin{array}{c}
         u_j^E  \\ v_j^E  
        \end{array} \right]_{(h,+)} =  c_{h , + } \: \left[ \begin{array}{c}
                                                              \tilde{c}_{R,L} ( E ) e^{   i k_p ( E ) (j - \ell) } \\
                                                             \tilde{t}_{ R,L} ( E ) e^{ - i k_h ( E ) ( j - \ell )  } 
                                                            \end{array} \right]
                                                            \;\;\; , \;\; ({\rm for} \: j \geq \ell + 1 )
\;\;\;\; ; 
\label{addiz.6}
\eneq
\noindent

 \item {\bf Incoming hole  from the right-hand lead}

 The corresponding eigenfunction is given by  
\beq
 \left[ \begin{array}{c}
         u_j^E  \\ v_j^E  
        \end{array} \right]_{(h,-)} =  c_{h , -  } \: \left[ \begin{array}{c}
                                                              \tilde{c}_{L,R} ( E ) e^{   - i k_p ( E ) ( j - 1)  } \\
                                                             \tilde{t} _{ L,R} ( E ) e^{  i k_h ( E ) ( j - 1 )  } 
                                                             \end{array} \right] \;\;\; , \;\; ({\rm for} \: j \leq 0 )
\:\:\:\: , 
\label{addiz.7}
\eneq
\noindent
and 
\beq
 \left[ \begin{array}{c}
         u_j^E  \\ v_j^E  
        \end{array} \right]_{(h,-)} =  c_{h , - } \: \left[ \begin{array}{c}
                                                            \tilde{a}_{R,R} ( E ) e^{  i k_p ( E ) ( j - \ell )  } \\
                                                           e^{   i k_h ( E ) ( j - \ell )  } +    \tilde{r}_{ R,R} ( E ) e^{ -  i k_h ( E ) ( j - \ell )  } 
                                                            \end{array} \right]
                                                            \;\;\; , \;\; ({\rm for} \: j \geq \ell + 1 )
\;\;\;\; . 
\label{addiz.8}
\eneq
\noindent
 \end{itemize}
As a general remark, we note that, among the possible physical processes we list above, the  CAR  possesses 
the remarkable property of being  effective in creating highly 
entangled, distant particle-hole pairs \cite{beenakker_1}.  Notably, the CAR process
appears as a consequence of nonzero mass $\epsilon_d$ for the massive subgap modes and  it is accordingly expected to 
disappear in the SRTP, as we discuss in the main text.

  We now  review in detail the relation between 
the Green's function and the scattering amplitudes contained in the eigenfunctions listed above.

\subsection{Relation between the scattering amplitudes and the fully dressed Green's functions of the system}
\label{scat_green}

To spell out the relation between scattering amplitudes and Green's functions, we start by considering 
the Green's functions in Eq.(\ref{ap1.1}), which can be readily expressed in terms of the  solutions of the
BdG equations, by going through the 
representation of the fermion operators in real space in terms of the eigenmodes of the whole system 
Hamiltonian. This is determined by the equations   

\begin{eqnarray}
c_j &=& \sum_{E > 0 } \: \sum_{a} \{ [ u_j^E ]_a \Gamma_{ E , a } + [ v_j^E ]_a^* \Gamma_{E , a}^\dagger \}
\nonumber \\
c_j^\dagger &=& \sum_{E > 0 } \: \sum_{a} \{ [ v_j^E ]_a \Gamma_{ E , a } + [ u_j^E ]_a^* \Gamma_{E , a}^\dagger \}  
\;\;\;\; , 
\label{sgf.2}
\end{eqnarray}
\noindent
with $a \in \{ (p,+) , ( p , - ) , (h , + ) , ( h , - ) \}$ and $\{ \Gamma_{E , a } , \Gamma_{E , a}^\dagger \}$ being the corresponding 
energy eigenmodes satisfying the anticommutator algebra
\beq
\{ \Gamma_{E , a } , \Gamma_{E' , a' }^\dagger  \} = \delta_{E , E' } \delta_{a , a' }. 
\:\:\:\: . 
\label{sgf.3}
\eneq
\noindent
As a result, inserting Eqs.(\ref{sgf.3}) into Eqs.(\ref{ap1.1}) and
moving to the mixed (real space-frequency) representation, one obtains 
\beq
{\bf C}_{(X , X') ; (j , j' )} ( i \omega )  =  \sum_{ E > 0 } \:\sum_a 
 \frac{1}{ i \omega - E } 	\: \left[ \begin{array}{cc} ( u_{j}^E  )_a ( u_{j'}^E )_a^* & ( u_{j}^E )_a ( v_{j'}^E )_a^*  \\
 ( v_{j} ) _a^E ( u_{j'}^E )_a^* & ( v_{j} )_a^E ( v_{j'}^E )_a^*  \end{array} \right]  +  
  \sum_{ E > 0 } \:\sum_a \frac{1}{ i \omega + E } \: \left[ \begin{array}{cc} ( v_j^E)_a^* (  v_{j'}^E )_a&  ( v_j^E)_a^*   ( u_{j'}^E )_a \\ 
  ( u_j^E )_a^*  ( v_{j'}^E )_a & ( u_j^E )_a^* ( u_{j'}^E )_a \end{array} \right] 
\:\:\:\: ,
\label{supercond.6}
\eneq
\noindent
with, respectively, $ ( X , X' ) = (L , L )$ for $j , j' \leq 0$, $ (X , X') = (L , R )$ for $j\leq 0 , j'\geq \ell + 1$, 
$(X , X' ) = ( R , L )$ for $j \geq \ell + 1 , j' \leq 0$, and $( X , X' ) = ( R , R)$ for $j , j' \geq \ell + 1$.  

Now, to keep in touch with the results that we provide in  Eqs.(\ref{grs.6},\ref{grs.7},\ref{grs.8},\ref{grs.9}), we compute the Green's functions in terms of 
the solutions of the BdG  equations by expanding the momenta $k_p ( E ) , k_h ( E )$ around the Fermi momentum $k_F$, 
defined by $- 2 J \cos ( k_F ) - \mu = 0 $,   as 
\begin{eqnarray}
k_p ( E ) &=& k_F + \frac{E}{v} \nonumber \\
k_h ( E ) &=& k_F - \frac{E}{v} 
\:\:\:\: , 
\label{sgf.4}
\end{eqnarray}
\noindent
with the Fermi velocity $v = 2 J \sin ( k_F )$, which also corresponds to setting $\omega = 0$ in the function $v ( \omega )$, so 
that we substitute $v ( \pm \omega )$ with $v ( \omega = 0 ) = 2   i J \sin ( k_F ) = i v$. Without entering the details of 
a straightforward, though tedious, calculation, one  eventually obtains 
\begin{eqnarray}
{\bf C}_{ ( L , L ) ; ( j , j' ) } ( i \omega ) &=&  \frac{1}{i v} \:
\left[ \begin{array}{cc} e^{  i k_F| j - j ' | - \frac{\omega}{v} | j - j ' | }   & 0 \\ 0 & e^{ - i k_F| j - j ' |  - \frac{\omega}{v} | j - j ' | } 
\end{array} \right]  \, \, \, ,
\nonumber \\
&+& \frac{1}{i v} \:
\left[ \begin{array}{cc}  r_{L,L} (  i \omega ) 
e^{  - i k_F ( j + j' - 2 ) +  \frac{\omega}{v} ( j + j ' - 2 ) } & \tilde{a}_{L,L} (  i \omega ) e^{ - i k_F ( j - j' ) 
+     \frac{\omega}{v} ( j + j ' - 2 ) } \\  a_{L,L} ( i \omega ) e^{     i k_F ( j - j' ) 
+     \frac{\omega}{v} ( j + j ' - 2 ) } &   \tilde{r}_{L,L} (  i \omega ) 
e^{   i k_F ( j + j' - 2 ) +  \frac{\omega}{v} ( j + j ' - 2 ) }  
\end{array} \right]  \, \, \, ,
\label{ls.4bis}
\end{eqnarray}
\noindent
\beq
{\bf C}_{ ( L , R ) ; ( j , j' ) } ( i \omega ) =  \frac{1}{iv} \:
\left[ \begin{array}{cc} t_{L,R}  (  i \omega ) 
e^{   - i k_F ( j - j' +  \ell - 1 ) +  \frac{ \omega}{v} ( j - j ' +  \ell - 1  ) } & \tilde{ c}_{L,R} (  i \omega ) e^{  - i k_F ( j + j'  - \ell - 1) 
+      \frac{ \omega}{v} ( j - j ' +  \ell - 1  )  } \\  c_{L,R} (  i \omega ) e^{ i k_F ( j + j'  - \ell - 1) 
+       \frac{ \omega}{v} ( j - j ' +  \ell - 1  )  } &  
  \tilde{t}_{L,R} (  i \omega ) 
e^{  i k_F ( j -j' +  \ell - 1 ) +       \frac{ \omega}{v} ( j - j ' +  \ell - 1  )}  
\end{array} \right] \, \, \, , 
\label{ls.6bis}
\eneq
\noindent
\beq
{\bf C}_{ ( R , L ) ; ( j , j' ) } ( i \omega ) =  \frac{1}{i v} \:
\left[ \begin{array}{cc} t_{R,L}  (  i \omega ) 
e^{   i k_F ( j - j' -  \ell + 1 ) -  \frac{ \omega}{v} ( j - j ' -  \ell + 1  ) } & \tilde{c}_{R,L} (  i \omega ) e^{  i k_F ( j + j'  - \ell - 1) 
-     \frac{ \omega}{v} ( j - j ' -  \ell +1  )  } \\  c_{R,L} (  i \omega ) e^{  - i k_F ( j + j'  - \ell - 1) 
-      \frac{ \omega}{v} ( j - j ' - \ell + 1  )  } &  
  \tilde{t}_{R,L} (  i \omega ) 
e^{ - i k_F ( j -j' -  \ell + 1 ) +       \frac{ \omega}{v} ( j - j ' -  \ell + 1  )}  
\end{array} \right]
\:\:\:\: .
\label{ls.7bis}
\eneq
\noindent
and, finally
\begin{eqnarray}
{\bf C}_{ ( R , R ) ; ( j , j' ) } ( i \omega ) &=&  \frac{1}{i v} \:
\left[ \begin{array}{cc} e^{    i k_F | j - j ' |  - \frac{\omega}{v} | j - j ' |  }  & 0 \\  0 & e^{ - i k_F | j - j ' |  - \frac{\omega}{v} | j - j ' | }  
\end{array} \right]
\nonumber \\
 &+&  \frac{1}{i v} \:
\left[ \begin{array}{cc}   r_{R,R}  (  i \omega ) 
e^{   i k_F ( j + j' - 2  \ell ) -  \frac{ \omega}{v} ( j + j ' - 2 \ell  ) } & \tilde{a}_{R,R} (  i \omega ) e^{   i k_F ( j - j' ) 
-     \frac{\omega}{v} ( j + j ' - 2  \ell) } \\  a_{R,R} (  i \omega ) e^{ -  i k_F ( j - j' ) 
-     \frac{\omega}{v} ( j + j ' - 2 \ell  ) } &   \tilde{r}_{ R, R} (  i \omega ) 
e^{  -  i k_F ( j + j' - 2 \ell ) -  \frac{\omega}{v} ( j + j ' - 2 \ell ) }  
\end{array} \right]
\:\:\:\: .
\label{ls.5bis}
\end{eqnarray}
\noindent
By direct comparison, one therefore obtains 
\beq
\left[ \begin{array}{cccc}
 r_{L, L}( i \omega )  & \tilde{a}_{L,L} ( i \omega )  &  t_{L, R}( i \omega )  & \tilde{c}_{L,R} ( i \omega )  \\
 a_{L , L } ( i \omega ) & \tilde{r}_{L,L} ( i \omega )  & c_{L , R } ( i \omega ) & \tilde{t}_{L,R} ( i \omega ) \\
t_{R, L}( i \omega )  & \tilde{c}_{R,L} ( i \omega ) & r_{R, R}( i \omega )  & \tilde{a}_{R,R} ( i \omega ) \\
c_{R , L } ( i \omega ) & \tilde{t}_{R,L} ( i \omega ) & a_{R , R } ( i \omega ) & \tilde{r}_{R,R} ( i \omega ) 
       \end{array} \right] = \left[ \begin{array}{cccc}  - 1 + \frac{  v \Upsilon_{L,L}^{(1,1)} ( i \omega ) }{i J^2} & 
- \frac{   v \Upsilon_{L,L}^{(1,2)} ( i \omega ) }{i J^2} &   \frac{  v \Upsilon_{L,R}^{(1,1)} ( i \omega ) }{i J^2} & - \frac{   v \Upsilon_{L,R}^{(1,2)} ( i \omega ) }{i J^2} \\
 - \frac{   v \Upsilon_{L,L}^{(2,1)} ( i \omega ) }{i J^2} &  - 1 + \frac{  v \Upsilon_{L,L}^{(2,2)} ( i \omega ) }{ i J^2}  
& - \frac{   v \Upsilon_{L,R}^{(2,1)} ( i \omega ) }{i J^2} &   \frac{   v \Upsilon_{L,R}^{(2,2)} ( i \omega ) }{i J^2}  \\
 \frac{  v \Upsilon_{R,L}^{(1,1)} ( i \omega ) }{i J^2} & - \frac{   v \Upsilon_{R,L}^{(1,2)} ( i \omega ) }{i J^2} &
 - 1 + \frac{  v \Upsilon_{R,R}^{(1,1)} ( i \omega ) }{ i J^2} & - \frac{   v \Upsilon_{R,R}^{(1,2)} ( i \omega ) }{i J^2} \\
 - \frac{  v \Upsilon_{R,L}^{(2,1)} ( i \omega ) }{i  J^2} &   \frac{  v \Upsilon_{R,L}^{(2,2)} ( i \omega ) }{i  J^2}  & 
 - \frac{  v \Upsilon_{R,R}^{(2,1)} ( i \omega ) }{ i J^2} &  - 1 + \frac{   v \Upsilon_{R,R}^{(2,2)} ( i \omega ) }{ i J^2} 
\end{array}
\right]
\;\;\;\; . 
\label{relfin}
\eneq
\noindent
From Eqs.(\ref{relfin}), one therefore infers the set of relations between the 
scattering amplitudes and the Green's function of the central region, on which the matrices ${\bf \Upsilon}$ explicitly depend, 
via Eqs.(\ref{ultimate.3}). Therefore, one expects to be able to recover information about the central region dynamics by 
just measuring the scattering amplitudes or, more precisely, observable quantities directly related to the scattering amplitudes, 
such as the electric current, possibly in combination with current-current correlations (shot noise), as we discuss in 
the main text.

\section{The single-site simplified Hamiltonian}
\label{ssingle}

In this appendix, we   review the model Hamiltonian discussed in Ref.[\onlinecite{beenakker_1}], which, in the main text, we 
use as a guideline to discuss our results.  Following Ref.[\onlinecite{beenakker_1}],  we
consider a simplified Hamiltonian for the central region in the form 
\beq
{\cal H}_d = i \epsilon_d \gamma_L \gamma_R
\;\;\;\; ,
\label{es.1}
\eneq
\noindent
which, in terms of the operators  $d , d^\dagger$, defined as 
\begin{eqnarray}
 d &=& \frac{1}{2} [ \gamma_L + i \gamma_R ] \, \, ,
\nonumber \\
d^\dagger &=& \frac{1}{2} [ \gamma_L - i \gamma_R ]
\;\;\;\; , 
\label{e.2}
\end{eqnarray}
\noindent
can be rewritten as 
\beq
{\cal H}_d = 2 \epsilon_d  \left[ d^\dagger d - \frac{1}{2} \right]
\:\:\:\: . 
\label{es.3}
\eneq
\noindent
Assuming, as we have done in the main text, that the tunneling  strength is the same for the couplings to 
both the leads,  the tunneling Hamiltonian ${\cal H}_T$ reduces to  
\beq
{\cal H}_T = - t \{ c_{L,0}^\dagger - c_{L , 0 } \} \gamma_L - t \{ c_{R , \ell + 1}^\dagger - c_{R , \ell + 1 } \} \gamma_R
\:\:\:\: . 
\label{es.4}
\eneq
\noindent
We can regard the whole Hamiltonian ${\cal H}_d + {\cal H}_T$ as the effective Hamiltonian for a generic central region, of 
which we only consider the subgap mode. On explicitly expressing $\gamma_L , \gamma_R$ in terms of 
the (Dirac) eigenmodes $d , d^\dagger$, one obtains 
\beq
{\cal H_T} = - t [ c_{L , 0 }^\dagger  - c_{L , 0 } - i  ( c_{R , \ell + 1 }^\dagger - c_{R , \ell + 1 }  )  ] d 
- t d^\dagger [ c_{L ,  0 } - c_{L , 0 }^\dagger + i ( c_{ R , \ell + 1 } - c_{R , \ell + 1 }^\dagger  ) ] 
\:\:\:\: . 
\label{es.5a}
\eneq
\noindent
On comparing Eq.(\ref{es.5a}) with the general expression of ${\cal H}_T$ in Eq.(\ref{eql.7}) in terms of the eigenmodes of 
${\cal H}_d$, we therefore find (apart for an overall,  real multiplicative constant, which can be readily reabsorbed in the 
definition of $t$), that the wave function corresponding to the $d, d^\dagger$-modes is given by  
\begin{eqnarray}
&& u_1 = v_1 = 1 \nonumber \\
&& u_\ell = v_\ell = i 
\:\:\:\: . 
\label{es.6}
\end{eqnarray}
\noindent
In Nambu representation,  the relevant Green's functions of 
the central region are therefore given by  
\beq
{\bf G}_{d ; (1,1) } ( i \omega ) = {\bf G}_{d;(\ell , \ell )} ( i \omega ) = - \frac{2 i \omega }{ \omega^2 + \epsilon_d^2 }
\: \left[ \begin{array}{cc} 1 & 1 \\ 1 & 1 \ \end{array} \right]
\:\:\:\: , 
\label{ed.7}
\eneq
\noindent
and by
\beq
{\bf G}_{d; ( 1 , \ell ) } ( i \omega ) = - {\bf G}_{d ; ( \ell , 1 ) } ( i \omega ) =  - \frac{2 i \epsilon_d }{ \omega^2 + \epsilon_d^2 }
\: \left[ \begin{array}{cc} 1 & 1 \\ 1 & 1 \ \end{array} \right] 
\:\:\:\: . 
\label{es.8}
\eneq
\noindent
In order to perform a comparison with the results of Ref.[\onlinecite{beenakker_1}], we have to go through a further  low-energy approximation, in neglecting  the dependence 
on $\omega$ in ${\bf C}_{(L,L);(0,0)}^{(0)} ( i \omega )$, which yields 
\beq
{\bf \Upsilon}_{(L,L)} ( i \omega ) = \left[ \frac{1}{( J \omega - 4 t^2 \sin ( k_F ) )^2 + J^2 \epsilon_d^2 } \right] \: 
\left[ \begin{array}{cc}
- 2 i t^2 ( J \omega -4 t^2 \sin (k_F)) &     2 i t^2 ( J \omega -4 t^2 \sin (k_F)) \\     2 i t^2 ( J \omega -4 t^2 \sin (k_F)) & -     2 i t^2 ( J \omega -4 t^2 \sin (k_F))
       \end{array} \right]
\;\;\;\; , 
\label{es.9}
\eneq
\noindent
\beq
{\bf \Upsilon}_{ ( L,R)} ( i \omega ) =   \left[ \frac{1}{( J \omega - 4 t^2 \sin ( k_F ) )^2 + J^2 \epsilon_d^2 } \right] \: 
\left[ \begin{array}{cc} - 2 i J t^2 \epsilon_d &  2 i J t^2 \epsilon_d \\   2 i J t^2 \epsilon_d & - 2 i J t^2 \epsilon_d        
       \end{array} \right]
\;\;\;\; , 
\label{es.10}
\eneq
\noindent
\beq
{\bf \Upsilon}_{ ( R,L)} ( i \omega ) =  \left[ \frac{1}{( J \omega - 4 t^2 \sin ( k_F ) )^2 + J^2 \epsilon_d^2 } \right] \: 
\left[ \begin{array}{cc}   2 i J t^2 \epsilon_d &  - 2 i J t^2 \epsilon_d \\  - 2 i J t^2 \epsilon_d &  2 i J t^2 \epsilon_d        
       \end{array} \right]
\;\;\;\; , 
\label{es.11}
\eneq
\noindent
and, finally
\beq
{\bf \Upsilon}_{(R,R)} ( i \omega ) = \left[ \frac{1}{( J \omega - 4 t^2 \sin ( k_F ) )^2 + J^2 \epsilon_d^2 } \right] \: 
\left[ \begin{array}{cc} - 2 i t^2 ( J \omega -4 t^2 \sin (k_F)) & 2 i t^2 ( J \omega -4 t^2 \sin (k_F)) \\
      2 i t^2 ( J \omega -4 t^2 \sin (k_F)) & - 2 i t^2 ( J \omega -4 t^2 \sin (k_F))   
       \end{array} \right]
\;\;\;\; . 
\label{es.12}
\eneq
\noindent
Accordingly, the scattering amplitudes are given by 
\beq
\left[ 
\begin{array}{cc}
 r_{L,L} ( i \omega ) & \tilde{a}_{L,L} ( i \omega ) \\ a_{L,L} ( i \omega ) & \tilde{r}_{L,L} ( i \omega ) 
\end{array} \right] = \left[ 
\begin{array}{cc}
 r_{R,R} ( i \omega ) & \tilde{a}_{R,R} ( i \omega ) \\ a_{R,R} ( i \omega ) & \tilde{r}_{R,R} ( i \omega ) 
\end{array} \right] = 
\left[ \begin{array}{cc}
- 1 -\alpha ( i \omega) & -\alpha ( i \omega ) \\ -\alpha ( i \omega ) & - 1 - \alpha ( i \omega )         
       \end{array} \right]
\;\;\;\; , 
\label{es.13}
\eneq
\noindent
with 
\beq
\alpha ( i \omega ) = \frac{4 t^2 \sin ( k_F )  ( J \omega -4 t^2 \sin (k_F))  }{( J \omega - 4 t^2 \sin ( k_F ) )^2 + J^2 \epsilon_d^2}
\;\;\;\; ,
\label{es.14}
\eneq
\noindent
and 
\beq
\left[ \begin{array}{cc}
t_{L,R} ( i \omega ) & \tilde{c}_{L,R} ( i \omega ) \\ c_{L,R} ( i \omega ) & \tilde{t}_{L,R} ( i \omega )         
       \end{array} \right] = 
       - \left[ \begin{array}{cc}
t_{R,L} ( i \omega ) & \tilde{c}_{R,L} ( i \omega ) \\ c_{R,L} ( i \omega ) & \tilde{t}_{R,L} ( i \omega )         
       \end{array} \right] 
= \left[ \begin{array}{cc}
-\beta ( i\omega ) & -\beta ( i \omega ) \\ -\beta ( i \omega ) & -\beta  ( i \omega )           
         \end{array} \right] 
\;\;\;\; , 
\label{es.15}
\eneq
\noindent
being 
\beq 
\beta ( i \omega ) = \frac{4 t^2 \sin ( k_F ) J \epsilon_d }{( J \omega - 4 t^2 \sin ( k_F ) )^2 + J^2 \epsilon_d^2}
\;\;\;\; .
\label{es.16}
\eneq
\noindent
On back-rotating to real energies, we have to substitute $i \omega $ with $E + i 0^+$, which implies 
\begin{eqnarray}
\alpha ( i \omega ) &\to& \alpha ( E ) =- \frac{  \Gamma ( i E +   \Gamma ) }{ (  i E +  \Gamma )^2 +  \epsilon_d^2 } \nonumber \\
\beta ( i \omega ) &\to & \beta ( E ) =   \frac{\Gamma \epsilon_d}{  (i E +  \Gamma )^2 + \epsilon_d^2}
\;\;\;\; ,
\label{es.17}
\end{eqnarray}
\noindent
where we have set $\Gamma = 4 t^2 \sin ( k_F ) / J$. This is the basic result for the scattering amplitudes derived in 
Ref.[\onlinecite{beenakker_1}], which we used as a reference model Hamiltonian to discuss the subgap physics 
of our system.

\section{Derivation of the current and of the shot-noise in  the NSN junction}
\label{transport}

In this appendix, we review and discuss the derivation of the transport properties of the NSN junction. To do so, we rely on the Landauer-like 
scattering approach, in which one imagines to ''shoot in'' particles and holes against the SN interfaces from a thermal lead
at temperature $T  $: This allows us to recover the full expression of the current and of the zero-frequency shot-noise   just in terms of 
the single-particle and of the single hole scattering amplitudes. The approach we are going to review in the following is 
an adapted version of the general formalism developed and discussed in details in Ref.[\onlinecite{datta}].

Our starting point is 
the electric current operator over the link from $j$ to $j+1$, which  takes the form 
\beq
J_j = - i e J \{ c_j^\dagger c_{j+1} - c_{j+1}^\dagger c_j \}
\:\:\:\: . 
\label{cur.1}
\eneq
\noindent
Once expanding the single-fermion operators in the basis of the eigenstates obeying the scattering boundary conditions,  according 
to the Bogoliubov-Valatin transformations in Eqs.(\ref{sgf.2}), one obtains that the current operator in the Heisenberg representation 
takes the form 
 \begin{eqnarray}
J_j ( t ) &=& \sum_{a , a'  } \sum_{E , E' > 0 } \: \{ J_{(+,-);j}^{(a,a');(E,E')} e^{ i ( E - E' ) t } \Gamma_{a , E}^\dagger \Gamma_{a' , E' } 
+ J_{(-,+);j}^{(a,a');(E,E')} e^{- i ( E - E' ) t } \Gamma_{a , E}\Gamma_{a' , E' } ^\dagger \nonumber \\
&+&  J_{(+,+); j}^{(a,a');(E,E')} e^{ i ( E + E' ) t } \Gamma_{a , E}^\dagger \Gamma_{a' , E' }^\dagger  
+ J_{(-,-) ; j }^{(a,a');(E,E')} e^{- i ( E + E' ) t } \Gamma_{a , E}\Gamma_{a' , E' } \}
\:\:\:\: ,
\label{cur.3}
\end{eqnarray}
\noindent
with 
\begin{eqnarray}
J_{(+,-) ; j}^{(a,a');(E,E')} &=& - i e J \{ (u_j^E)_a^* (u_{j+1}^{E'} )_{a'} - ( u_{j+1}^E )_a^* ( u_j^{E'})_{a'} \} \nonumber \\
J_{(-,+) ; j}^{(a,a');(E,E')} &=& - i e J \{ (v_j^E)_a (v_{j+1}^{E'} )_{a'}^* - ( v_{j+1}^E )_a ( v_j^{E'})_{a'}^* \} \nonumber \\
J_{(+,+) ; j}^{(a,a');(E,E')} &=& - i e J \{ (u_j^E)_a^* (v_{j+1}^{E'} )_{a'} ^*- ( u_{j+1}^E )_a^* ( v_j^{E'})_{a'}^* \} \nonumber \\
J_{(- ,-) ; j}^{(a,a');(E,E')} &=& - i e J \{ (v_j^E)_a  (u_{j+1}^{E'} )_{a'} - ( v_{j+1}^E )_a   ( u_j^{E'})_{a'} \}
\;\;\;\; . 
\label{cur.4}
\end{eqnarray}
\noindent
Using Eqs.(\ref{cur.3}, \ref{cur.4}), in the following we compute the  current, as well as the zero-frequency shot noise.

\subsection{Current flowing across the normal leads}
\label{current}

Assuming that the two leads are biased at a voltage $V$ with respect to 
the central region,  basically means  that particles 
and/or holes  enter  the central region from thermal reservoirs, so that particles emerge
from the reservoirs at a chemical potential $\mu - e V$, $\mu$ being the equilibrium chemical potential 
of the whole NSN junction, while, accordingly, holes emerge at a chemical potential $\mu + e V$. In terms of 
average energy of the eigenmodes of the total Hamiltonian, this basically implies 
\begin{eqnarray}
&& \langle \Gamma_{a , E}^\dagger \Gamma_{a' , E' } \rangle = \delta_{a,a'} \delta_{E,E'} \: f ( E - e V )  \;\;  \,\ \, \, {\rm if}  \, \, \, \: a \in \{ (p,+), (p,-) \} \nonumber \\
&& \langle \Gamma_{a , E}^\dagger \Gamma_{a' , E' } \rangle = \delta_{a,a'} \delta_{E,E'} \: f ( E + e V ) \;\;  \, \, \, {\rm if} \, \, \, \,  \: a \in \{ (h,+), (h,-) \} \nonumber \\
&& \langle \Gamma_{a,E}^\dagger \Gamma_{a' , E'}^\dagger \rangle = \langle \Gamma_{a,E} \Gamma_{a' , E'} \rangle  = 0
\;\;\;\; ,
\label{cur.5}
\end{eqnarray}
\noindent
with $f  ( E )$ being the Fermi distribution function with chemical potential $\mu$. 
Taking  Eqs.(\ref{cur.5}) into account and considering the explicit form of the current operator,  one eventually obtains,
in the zero-temperature limit 

\begin{eqnarray}
 I_L &\to& \frac{2 e}{2 \pi} \: \int_0^{eV} \: d E \:  \{ |a_{L,L} ( E ) |^2 + | c_{L,R} ( E ) |^2 \}  \, \,\ \, , \nonumber \\
 I_R &\to& - \frac{2e}{2 \pi} \: \int_0^{eV} \: d E \:   \{ |a_{R,R} ( E ) |^2 + | c_{R,L} ( E ) |^2 \} 
 \:\:\:\: . 
 \label{cur.9}
\end{eqnarray}
\noindent
As a next step, we now compute  the zero-frequency shot noise.

\subsection{Zero-frequency shot noise}
\label{noise}

The zero-frequency shot noise  at voltage bias $V$ is defined from the Fourier transform of the real-time current-current correlation 
functions. In general, considering current operators in sites $j,j'$, one sets 
\beq
R_{j,j'} ( \Omega ) = \int_{-\infty}^\infty \: d t \: e^{ i \Omega t} \: \frac{1}{2} \: \langle \{  \delta J_j (t ) , \delta J_{j'  } ( 0 ) \} \rangle 
\:\:\:\: , 
\label{cur.10}
\eneq
\noindent
with 
\beq
\delta J_j ( t ) = J_j ( t ) - \langle J_j ( t ) \rangle
\:\:\:\: . 
\label{cur.11}
\eneq
\noindent
Therefore, the zero-frequency noise power is defined as 
\beq
{\cal P}_{j,j^{\prime}} = \lim_{\Omega \to 0 } R_{j,j'} ( \Omega ) 
\:\:\:\: . 
\label{np.1}
\eneq
\noindent
In the following, we are interested in computing ${\cal P}_{j , j'} ( 0 )$ in the case in which $j , j'$ both belong to 
the same lead, as well as in the case in which they belong to different leads. On implementing the formalism outlined above, 
after  long, though straightforward calculations, one obtains, in the zero-temperature limit  

\begin{eqnarray}
 {\cal P}_{L,L} ( 0 ) &=& \frac{e^2}{2 \pi} \; \int_0^{eV} \: d E \: \{ [ | r_{L,L} ( E ) |^2 + | t_{L,R} ( E ) |^2 ] [ 1 -  | r_{L,L} ( E ) |^2 - | t_{L,R} ( E ) |^2 ]
 \nonumber \\
 &+&  [ | a_{L,L} ( E ) |^2 + | c_{L,R} ( E ) |^2 ] [ 1 -  | a_{L,L} ( E ) |^2 - | c_{L,R} ( E ) |^2 ] \nonumber \\
 &+& 2 | ( r_{L,L} (E) )^* a_{L,L} ( E ) +   ( t_{L,R} (E) )^* c_{L,R} ( E ) |^2 \} 
 \:\:\:\:, 
 \label{cur.17}
\end{eqnarray}
\noindent
as well as
\begin{eqnarray}
 {\cal P}_{R,R} ( 0 ) &=& \frac{e^2}{2 \pi} \; \int_0^{eV} \: d E \: \{ [ | r_{R,R} ( E ) |^2 + | t_{R,L} ( E ) |^2 ] [ 1 -  | r_{R,R} ( E ) |^2 - | t_{R,L} ( E ) |^2 ]
 \nonumber \\
 &+&  [ | a_{R,R} ( E ) |^2 + | c_{R,L} ( E ) |^2 ] [ 1 -  | a_{R,R} ( E ) |^2 - | c_{R,L} ( E ) |^2 ] \nonumber \\
 &+& 2 | ( r_{R,R} (E) )^* a_{R,R} ( E ) +   ( t_{R,L} (E) )^* c_{R,L} ( E ) |^2 \} 
 \:\:\:\:.
 \label{cur.17bis}
\end{eqnarray}
\noindent
On the contrary, if $j$ and $j'$ belong to two different leads, one obtains 
 \begin{eqnarray}
  {\cal P}_{L , R} ( 0 ) &=&    \frac{e^2}{2 \pi} \; \int_0^{eV} \: d E \: \{ |  ( r_{L,L} ( E ))^* t_{R,L} ( E ) + (t_{L,R} ( E ))^* r_{R,R} ( E ) |^2 +
  | ( a_{L,L} ( E ) )^* c_{R,L} ( E ) + ( c_{L,R} ( E ) )^* a_{R,R} ( E ) |^2 \nonumber \\
  &-& | (r_{L,L} ( E ))^* c_{R,L} ( E ) + (t_{L,R} ( E ))^* a_{R,R} ( E ) |^2 - |  ( a_{L,L} ( E ) )^* t_{L,R} ( E ) + ( c_{L,R} ( E ) )^* 
 r_{R,R} ( E ) |^2 \}
 \:\:\:\: , 
 \label{cur.19}
\end{eqnarray}
\noindent
as well as 
\begin{eqnarray}
  {\cal P}_{R , L} ( 0 ) &=&    \frac{e^2}{2 \pi} \; \int_0^{eV} \: d E \: \{ |  ( r_{R,R} ( E ))^* t_{L,R} ( E ) + (t_{R,L} ( E ))^* r_{L,L} ( E ) |^2 +
  | ( a_{R,R} ( E ) )^* c_{L,R} ( E ) + ( c_{R,L} ( E ) )^* a_{L,L} ( E ) |^2 \nonumber \\
  &-& | (r_{R,R} ( E ))^* c_{L,R} ( E ) + (t_{R,L} ( E ))^* a_{L,L} ( E ) |^2 - |  ( a_{R,R} ( E ) )^* t_{R,L} ( E ) + ( c_{R,L} ( E ) )^* 
 r_{L,L} ( E ) |^2 \}
 \:\:\:\: . 
 \label{cur.19bis}
\end{eqnarray}
\noindent
The formulas in Eqs. (\ref{cur.17},\ref{cur.17bis},\ref{cur.19},\ref{cur.19bis}) have been used in the main text to compute the current 
and the noise power in the various phases of the LRK.

\bibliography{paper_bib_rev.bib}

\end{document}